\documentclass[a4paper,11pt]{article}
\usepackage{graphicx,epsfig}
\usepackage{fancyhdr,fancybox,float}
\usepackage[title]{appendix}
\usepackage{indentfirst}
\usepackage{verbatim}
\usepackage[sort&compress, numbers]{natbib}
\usepackage{geometry}
\usepackage{extarrows,chemarrow,xypic}
\usepackage{color}
\usepackage[footnotesize]{caption2}
\usepackage{microtype}
\usepackage{xcolor}
\DisableLigatures[f]{encoding = *, family = *}

\usepackage{times}

\usepackage{amssymb}
\usepackage{amsthm}
\usepackage{mathrsfs}
\usepackage{bbm}
\usepackage{hyperref}

\newcommand{\paperfont}{\fontsize{10pt}{1.1\baselineskip}\selectfont}
\definecolor{blue}{RGB}{0,50,200}

\begin{document}

\theoremstyle{definition}
\makeatletter
\thm@headfont{\bf}
\makeatother
\newtheorem{definition}{Definition}
\newtheorem{example}{Example}
\newtheorem{theorem}{Theorem}
\newtheorem{lemma}{Lemma}
\newtheorem{corollary}{Corollary}
\newtheorem{remark}{Remark}
\newtheorem{proposition}{Proposition}


\lhead{}
\rhead{}
\lfoot{}
\rfoot{}

\renewcommand{\refname}{References}
\renewcommand{\figurename}{Figure}
\renewcommand{\tablename}{Table}
\renewcommand{\proofname}{Proof}

\newcommand{\diag}{\mathrm{diag}}
\newcommand{\one}{\mathbbm{1}}
\newcommand{\Pnum}{\mathbb{P}}
\newcommand{\Rnum}{\mathbb{R}}
\newcommand{\Enum}{\mathbb{E}}
\newcommand{\set}[1]{\left\{#1\right\}}
\newcommand{\style}{\setlength{\itemsep}{1pt}\setlength{\parsep}{1pt}\setlength{\parskip}{1pt}}

\title{\textbf{Exact power spectrum in a minimal hybrid model of stochastic gene expression oscillations}}
\author{Chen Jia$^{1,*}$,\;\;\;Hong Qian$^2$,\;\;\;Michael Q. Zhang$^{3}$\\
\footnotesize $^1$ Applied and Computational Mathematics Division, Beijing Computational Science Research Center, Beijing 100193, China. \\
\footnotesize $^3$ Department of Applied Mathematics, University of Washington, Seattle, WA 98195, U.S.A. \\
\footnotesize $^2$ Department of Biological Sciences, University of Texas at Dallas, Richardson, TX 75080, U.S.A. \\
\footnotesize $^*$ Correspondence: chenjia@csrc.ac.cn}
\date{}
\maketitle
\thispagestyle{empty}

\paperfont

{\abstract
Stochastic oscillations in individual cells are usually characterized by a non-monotonic power spectrum with an oscillatory autocorrelation function. Here we develop an analytical approach of stochastic oscillations in a minimal hybrid model of stochastic gene expression including promoter state switching, protein synthesis and degradation, as well as a genetic feedback loop. The oscillations observed in our model are noise-induced since the deterministic theory predicts stable fixed points. The autocorrelated function, power spectrum, and steady-state distribution of protein concentration fluctuations are computed in closed form without making any approximations. Using the exactly solvable model, we illustrate sustained oscillations as a circular motion along a stochastic hysteresis loop induced by gene state switching. A triphasic stochastic bifurcation upon the increasing strength of negative feedback is observed, which reveals how stochastic bursts evolve into stochastic oscillations. In our model, oscillations tend to occur when the protein is relatively stable and when gene switching is relatively slow. Translational bursting is found to enhance the robustness and broaden the region of stochastic oscillations. These results provide deeper insights into R. Thomas' two conjectures for single-cell gene expression kinetics.}

\section{Introduction}
The engineering concept of feedback loops has been extremely useful in understanding nonlinear biological dynamics \cite{murray2007mathematical} and cellular regulation \cite{alon2006introduction} in terms of biochemical kinetics \cite{qian2012cooperativity}. In connection to gene regulatory networks,  Ren\'{e} Thomas proposed two conjectures in $1981$ \cite{thomas1981relation}: (i) The presence of a positive feedback loop is a necessary condition for multiple stable states; (ii) The presence of a negative feedback loop is a necessary condition for sustained oscillations. Many efforts since then have been devoted to the investigation and proof of these two conjectures. However, the majority of previous studies are based on the theory of deterministic (continuous and discrete) dynamical systems, such as ordinary differential equations (ODEs) and Boolean networks. In the context of continuous dynamical systems, both the first and second conjectures have been extensively studied based on the concept of local or global interaction graphs \cite{plahte1995feedback, gouze1998positive, snoussi1998necessary, cinquin2002positive, soule2003graphic, richard2011stable}.

Over the past two decades, large amounts of single-cell experiments, some with single-molecule sensitivity, have revealed a large cell-to-cell variation in the numbers of mRNA and protein molecules in isogenic populations due to stochasticity in gene expression and the low copy numbers of many components, including DNA and important regulatory molecules \cite{elowitz2002stochastic, golding2005real, cai2006stochastic, raj2008nature, taniguchi2010quantifying}. To explain noisy experimental data, notable breakthroughs have been made in the kinetic theory of single-cell stochastic gene expression \cite{paulsson2005models, schnoerr2017approximation}. The stochastic properties of gene regulation have been explored mostly by stochastic simulations and to a lesser extent by analytical solutions of various discrete, continuous, and hybrid gene expression models. Discrete models are those in which the numbers of genes, mRNAs, and proteins change by discrete integer amounts when reactions occur \cite{shahrezaei2008analytical}. In continuous models, fluctuations correspond to hops on the real axis rather than on the integer axis \cite{friedman2006linking, mackey2013dynamic, bokes2015protein, jkedrak2016time, chen2020limit}. In hybrid models, the fluctuations of genes are modeled discretely, while the fluctuations of mRNAs and proteins are modeled in a continuous sense \cite{karmakar2004graded, raj2006stochastic, dattani2017stochastic, bressloff2017stochastic, lin2018efficient, jia2019single}. Here we focus on the hybrid model since (i) it serves as an accurate approximation of the discrete model when protein numbers are relatively large \cite{jia2017emergent, wang2023poisson}, and (ii) the protein abundance is usually modelled as a continuous variable in previous studies on oscillations.

Sustained oscillations are common in nature. Most previous studies about biological oscillations are based on deterministic modelling and analysis \cite{novak2008design, tsai2008robust, li2017incoherent}. Some of the best-understood biological oscillators include p53 responses \cite{lev2000generation, lahav2004dynamics, geva2006oscillations}, NF-kB responses \cite{hoffmann2002ikappab, nelson2004oscillations, tay2010single}, cell cycle \cite{li2004yeast, ferrell2011modeling}, circadian rhythm \cite{lee2000interconnected, gallego2007post}, calcium spikes \cite{meyer1988molecular, qi2020oscillation}, and various synthetic oscillators \cite{elowitz2000synthetic, stricker2008fast, barnes2011bayesian}. However, there are growing observations that gene expression in single cells can display stochastic oscillations, and noise can play a vital role in enhancing or weakening the robustness of these oscillations \cite{vellela2010}. Statistically, noise-induced oscillations are usually quantified by a non-monotonic power spectrum with an oscillatory autocorrelation function \cite{qian2000pumped, bratsun2005delay}. In recent years, significant progress has been made in the analytical theory of stochastic oscillations in biochemical reaction networks, where the autocorrelation function and power spectrum of concentration fluctuations of various biochemical species are analytically computed and analyzed \cite{bratsun2005delay, mckane2007amplified, dauxois2009enhanced, realpe2012demographic, thomas2012slow, toner2013effects, thomas2013signatures, thomas2014phenotypic, jia2021frequency, jia2022concentration, gupta2022frequency}. However, these studies are often based on various approximation methods, and hence are not exact. Common approximations include the linear noise approximation (LNA) \cite{mckane2007amplified, dauxois2009enhanced, realpe2012demographic, thomas2012slow, toner2013effects}, corrected LNA \cite{thomas2013signatures}, large delay approximation \cite{bratsun2005delay}, and Pad\'{e} approximation \cite{gupta2022frequency}. The LNA works well in the macroscopic limit of large system size. It approximates the discrete Markovian model of a biochemical network by an Ornstein-Uhlenbeck process, whose probability distribution is Gaussian. Using the LNA, the power spectrum can be approximately computed for any monostable biochemical system near a Hopf bifurcation \cite{mckane2007amplified}. The LNA has also been improved to include higher-order corrections to the spectrum and it turns out that the corrected LNA is valid over a wider range of system size \cite{thomas2013signatures}. The large delay approximation works well in non-Markovian biochemical networks with time delay. It assumes that the delay is large compared to other characteristic times of the system so that the events before and after delay are effectively decoupled \cite{bratsun2005delay}. The Pad\'{e} approximation is applicable to general nonlinear biochemical network. It approximates the power spectrum by a rational function. However, the coefficients of the rational function have to be computed approximately by stochastic simulations and hence the method is only semi-analytical. Thus far, there is still a lack of an exact theory of stochastic gene expression oscillations at the single-cell level.

In this paper, we develop an analytical approach and present a comprehensive mathematical analysis for stochastic oscillations in a minimal stochastic gene expression model with gene state switching, protein synthesis, protein degradation, and a genetic feedback loop. The structure of the paper is as follows. In Sec. 2, we describe our hybrid ODE model and obtain the deterministic rate equation of the system. In particular, we show that the deterministic theory only predicts stable fixed points and thus oscillations can only be observed when noise is present. In Sec. 3, we derive the exact analytical expressions of the autocorrelation function, power spectrum, and steady-state distribution of protein concentration fluctuations. In Sec. 4, based on the analytical results, we show that sustained oscillations fail to be observed when the gene is unregulated or positively regulated. In a negative feedback loop, we reveal the emergence of a triphasic stochastic bifurcation as the negative feedback strength increases. Furthermore, we provide an intuitive and mechanistic picture of Thomas' two conjectures for single-cell gene expression. With this picture, sustained oscillations are shown to be a circular motion along a stochastic hysteresis loop induced by gene state switching. In Sec. 5, we analytically compute the autocorrelation function and power spectrum in a more complex hybrid model, where translational bursting is taken into account. We also investigate the influence of burstiness on oscillations and find that protein bursting enhances the robustness and broaden the region of stochastic oscillations. We conclude in Sec. 6.

\section{Model}

\subsection{Stochastic hybrid model}
\begin{figure}[!htb]
\centerline{\includegraphics[width=1.0\textwidth]{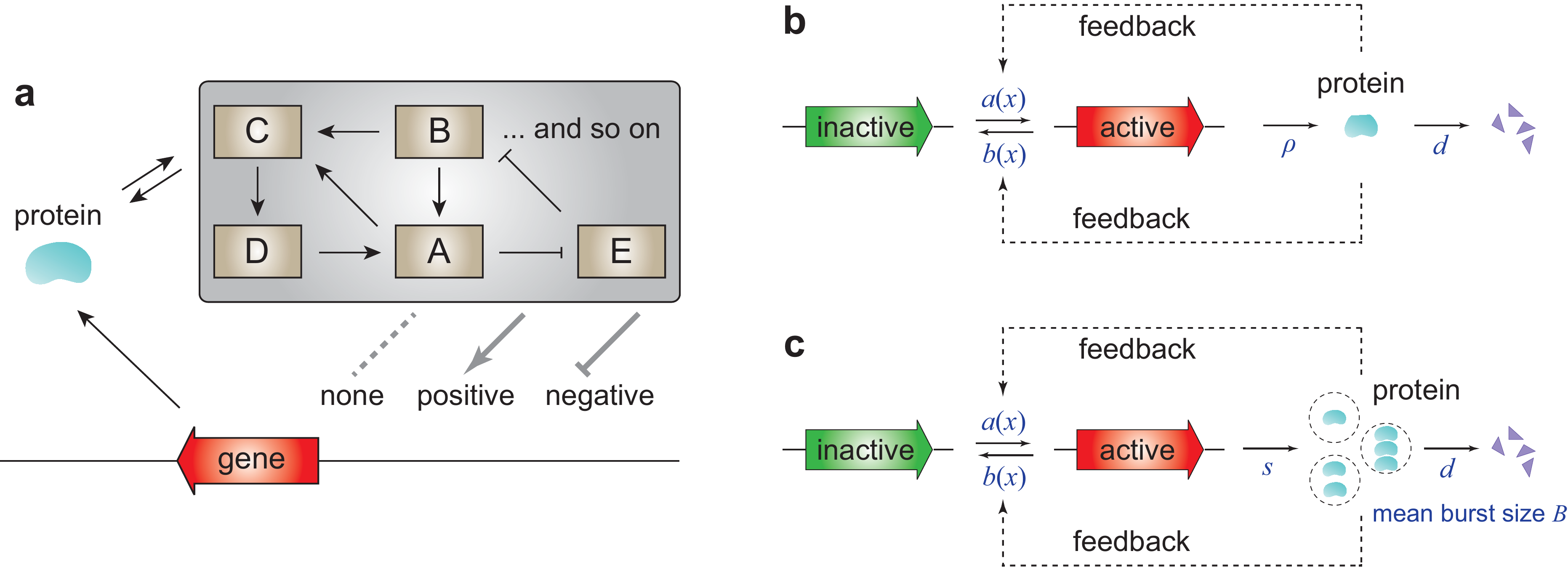}}
\caption{\textbf{Schematics of stochastic gene expression in living cells.}
\textbf{(a)} Three types of gross feedback topologies. Gene regulatory networks in a living cell can be extremely complex, involving numerous feedback loops and signaling steps (grey box). If we focus on a particular gene of interest (red), then there are three types of fundamental regulatory relations: no feedback (none), positive feedback, and negative feedback. The dotted line represents that there is no link between adjacent nodes.
\textbf{(b)} Two-state model of gene expression involving gene state switching, protein synthesis, and protein degradation. Due to feedback regulation, the switching rates between the active and inactive gene states, $a(x)$ and $b(x)$, depend on the protein concentration $x$. Here protein synthesis occurs in a non-bursty manner.
\textbf{(c)} Two-state model of gene expression involving gene state switching, protein synthesis, and protein degradation. Here protein synthesis occurs in a bursty manner.}\label{network}
\end{figure}

Gene regulatory networks can be tremendously complex, involving numerous signaling steps and genetic feedback loops. However, the situation becomes much simpler if we focus on a particular gene and the feedback loop regulating it \cite{lestas2010fundamental}. In general, there are three types of gross feedback topologies for the gene of interest: no feedback, positive feedback, and negative feedback (Fig. \ref{network}(a)). Here we consider a minimal stochastic gene expression model that can produce sustained oscillations. The model includes synthesis and degradation of protein, as well as switching of the gene between an active and an inactive state (Fig. \ref{network}(b)). The evolution of the protein concentration $x$ is modeled by a hybrid ODE \cite{karmakar2004graded, raj2006stochastic, dattani2017stochastic, bressloff2017stochastic, lin2018efficient, jia2019single}, also called a piecewise deterministic Markov process:
\begin{equation*}
\xymatrix@R = 0.5cm{
\dot{x} = \rho-dx \ar@<0.2ex>@^{->}[d]^{b(x)} & (\textrm{active state}),\\
\dot{x} = -dx \ar@<0.2ex>@^{->}[u]^{a(x)} & (\textrm{inactive state}),}
\end{equation*}
where $\rho$ is the synthesis rate of protein when the gene is active, and $d$ is the degradation rate of protein. Protein synthesis does not occur in the inactive gene state. Due to feedback regulation, the protein concentration $x$ will directly or indirectly affect the switching rates $a(x)$ and $b(x)$ of the gene between the two states. There are two reasons why we use a hybrid ODE to model gene expression oscillations. First, the hybrid model has been shown to be an accurate approximation of the discrete chemical master equation (CME) model when protein numbers are relatively large \cite{jia2017emergent, jia2019single}. Second, the protein abundance is usually modelled as a continuous variable in previous studies on oscillations.

If the gene is unregulated, then the gene switching rates $a(x) = a$ and $b(x) = b$ are independent of $x$. In a positive (negative) feedback network, the gene activation rate $a(x)$ is an increasing (decreasing) function of $x$ due to epigenetic controls such as association (dissociation) of activators, while the gene inactivation rate $b(x)$ is an decreasing (increasing) function of $x$ due to epigenetic controls such as dissociation (association) of repressors. To establish an analytical theory of stochastic oscillations, we assume that $a(x)$ and $b(x)$ linearly depend on $x$:
\begin{equation}\label{functional}
a(x) = a-ux,\;\;\;b(x) = b+ux,
\end{equation}
where $|u|$ characterizes the strength of feedback regulation with $u = 0$ corresponding to unregulated genes, $u>0$ corresponding to negatively regulated genes, and $u<0$ corresponding to positively regulated genes. The functional forms of $a(x)$ and $b(x)$ introduced above are essentially consistent with a popular model for NF-$\kappa$B nuclear translocation, a most well-understood biological oscillator \cite{ashall2009pulsatile}. Such linear dependence of $a(x)$ and $b(x)$ on $x$ has been widely used in the study of autoregulatory gene circuits \cite{hornos2005self, grima2012steady, kumar2014exact, jia2020small, jia2020dynamical}. However, most previous studies allow only one of $a(x)$ and $b(x)$ to be dependent on $x$, while the other is a constant. Here we require both $a(x)$ and $b(x)$ to be dependent on $x$.

The microstate of the gene can be represented by an ordered pair $(i,x)$, where $i$ is the gene state with $i = 0,1$ corresponding to the inactive and active states, respectively. Let $p_i(x,t)$ denote the probability density of the protein concentration at time $t$ when the gene is in state $i$. Then the evolution of the hybrid ODE model is governed by the Kolmogorov forward equation
\begin{equation}\label{forwardode}\left\{
\begin{split}
\partial_tp_0 &= d\partial_x(xp_0)+b(x)p_1-a(x)p_0,\\
\partial_tp_1 &= d\partial_x(xp_1)-\rho\partial_xp_1+a(x)p_0-b(x)p_1.
\end{split}\right.
\end{equation}

\subsection{Deterministic rate equation fails to produce sustained oscillations}
We first derive the deterministic rate equation for the hybrid model. Let $g(t) = \int_0^\infty p_1(x,t)dx$ denote the mean number of genes in the active state and let $m(t) = \Enum x(t) = \int_0^\infty xp(x,t)dx$ denote the mean protein concentration, where $x(t)$ is the protein concentration in a single cell at time $t$ and $p(x,t) = p_0(x,t)+p_1(x,t)$ is the probability density of the protein concentration. Integrating the second identity in Eq. \eqref{forwardode}, we obtain
\begin{equation*}
\begin{split}
\frac{d}{dt}g(t) &= \int_0^\infty\partial_tp_1dx = \int_0^\infty[(a-ux)p_0-(b+ux)p_1]dx \\
&= a[1-g(t)]-bg(t)-um(t) \\
&= a-(a+b)g(t)-um(t).
\end{split}
\end{equation*}
Similarly, using Eq. \eqref{forwardode} and integration by parts, we obtain
\begin{equation*}
\begin{split}
\frac{d}{dt}m(t) &= \int_0^\infty x\partial_tpdx
= \int_0^\infty x[d\partial_x(xp)-\rho\partial_xp_1]dx \\
&= \int_0^\infty(\rho p_1-dxp)dx \\
&= \rho g(t)-dm(t).
\end{split}
\end{equation*}
Hence $g(t)$ and $m(t)$ satisfy the following coupled set of ODEs:
\begin{equation}\label{ODE}
\frac{d}{dt}\begin{pmatrix}g(t) \\ m(t)\end{pmatrix}
= -T\begin{pmatrix}g(t) \\ m(t)\end{pmatrix}+\begin{pmatrix}a \\ 0\end{pmatrix},
\end{equation}
where
\begin{equation*}
T = \begin{pmatrix}
a+b & u\\-\rho & d
\end{pmatrix}
\end{equation*}
is a $2\times 2$ matrix. The two eigenvalues $\lambda_1$ and $\lambda_2$ of the matrix $T$ are the solutions to the quadratic equation
\begin{equation*}
\lambda^2-(d+a+b)\lambda+d(a+b)+\rho u = 0,
\end{equation*}
which has a discriminant $\Delta = (d-a-b)^2-4\rho u$. We assume that the gene network can reach a steady state, which implies that both $\lambda_1$ and $\lambda_2$ have positive real parts. Clearly, if the gene is unregulated or positively regulated, then $u\leq 0$ and thus $\Delta>0$, which shows that the two eigenvalues are both positive real numbers. However, for negative feedback loops ($u>0$), the discriminant $\Delta$ may be negative, which means that the two eigenvalues may become conjugate complex numbers.

We emphasize that Eq. \eqref{ODE} can be viewed as the deterministic rate equation of the stochastic hybrid model since it characterizes the evolution of the mean number of active genes and mean protein concentration. Note that it is a linear system of ODEs and hence it has a globally stable fixed point (since the two eigenvalues of $T$ have positive real parts). The fixed point is a node when the two eigenvalues are real and it is a focus when the two eigenvalues are conjugate complex numbers. This suggests that it is possible for the deterministic rate equation to display damped oscillations (when the fixed point is a focus), but the deterministic theory predicts that sustained oscillations can never occur.

\section{Exact solution}

\subsection{Exact solutions of the autocorrelation function and power spectrum}
We next focus on the oscillatory behavior of the hybrid model. Stochastic gene expression oscillations are usually characterized by the autocorrelation function and power spectrum. The former $C(t) = \textrm{Cov}_{ss}(x(0),x(t))$ is defined as the steady-state covariance of $x(0)$ and $x(t)$, and the latter $G(\omega) = \int_{-\infty}^\infty C(|t|)e^{-2\pi i\omega t}dt$ is defined as the Fourier transform of $C(|t|)$. In general, sustained oscillations fails to be observed if $G(\omega)$ is monotonically decreasing over $[0,\infty)$, while a non-monotonic $G(\omega)$ over $[0,\infty)$ serves as a characteristic signal of robust stochastic oscillations with the maximum point being the dominant frequency.

We next compute the analytical expressions of $C(t)$ and $G(\omega)$. To this end, let $r(t) = \int_0^\infty xp_1(x,t)dx$ denote the unnormalized mean of the protein concentration when the gene is active and let $s(t) = \int_0^\infty x^2p(x,t)dx$ denote the second moment of the protein concentration. Integrating Eq. \eqref{forwardode}, it is easy to check that $r(t)$ and $s(t)$ satisfy the following set of ODEs:
\begin{equation}\label{ODE2}
\frac{d}{dt}\begin{pmatrix}r(t) \\ s(t)\end{pmatrix}
= -R\begin{pmatrix}r(t) \\ s(t)\end{pmatrix}
+\begin{pmatrix}\rho g(t)+am(t) \\ 0\end{pmatrix},
\end{equation}
where
\begin{equation}\label{Rmatrix}
R = \begin{pmatrix}
d+a+b & u\\-2\rho & 2d
\end{pmatrix}
\end{equation}
is a $2\times 2$ matrix. In steady state, it follows from Eq. \eqref{ODE} that
\begin{equation}\label{ssm}
\begin{pmatrix}g(\infty) \\ m(\infty)\end{pmatrix}
= T^{-1}\begin{pmatrix}a \\ 0\end{pmatrix}
= \frac{a}{d(a+b)+\rho u}\begin{pmatrix}d \\ \rho\end{pmatrix},
\end{equation}
and it follows from Eqs. \eqref{ODE2} and \eqref{ssm} that
\begin{equation}\label{sss}
\begin{split}
\begin{pmatrix}r(\infty) \\ s(\infty)\end{pmatrix}
&= R^{-1}\begin{pmatrix}\rho g(\infty)+am(\infty) \\ 0\end{pmatrix}\\
&= \frac{\rho a(a+d)}{[d(a+b)+\rho u][d(d+a+b)+\rho u]}\begin{pmatrix}d \\ \rho\end{pmatrix}.
\end{split}
\end{equation}

To proceed, let $\alpha(t)$ denote the gene state in a single cell at time $t$. Note that $\alpha(t)$ and $x(t)$ are both stochastic processes. Since Eq. \eqref{ODE} is a linear system of ODEs, it can be solved exactly as
\begin{equation}\label{solution}
\begin{pmatrix}g(t) \\ m(t)\end{pmatrix}
= e^{-Tt}\begin{pmatrix}g(0) \\ m(0)\end{pmatrix}
+(I-e^{-Tt})T^{-1}\begin{pmatrix}a \\ 0\end{pmatrix}.
\end{equation}
Note that Eq. \eqref{solution} holds for any initial conditions. Let $m_i(x,t) = \Enum[x(t)|\alpha(0)=i,x(0)=x]$ denote the conditional mean of the protein concentration given that the initial gene state is $i$ and the initial protein concentration is $x$. It then follows from Eq. \eqref{solution} that
\begin{equation}\label{conditional}
\begin{pmatrix}g(t) \\ m_i(x,t)\end{pmatrix}
= e^{-Tt}\begin{pmatrix}i \\ x\end{pmatrix}
+(I-e^{-Tt})T^{-1}\begin{pmatrix}a \\ 0\end{pmatrix}.
\end{equation}
This clearly shows that
\begin{equation*}
\Enum[x(t)|\alpha(0),x(0)] = (0,1)e^{-Tt}\begin{pmatrix}\alpha(0) \\ x(0)\end{pmatrix}
+(0,1)(I-e^{-Tt})T^{-1}\begin{pmatrix}a \\ 0\end{pmatrix}.
\end{equation*}
Hence in steady state, we have
\begin{equation*}
\begin{split}
\Enum x(0)x(t) &= \Enum x(0)\Enum[x(t)|\alpha(0),x(0)] \\
&= \Enum x(0)\left[(0,1)e^{-Tt}\begin{pmatrix}\alpha(0) \\ x(0)\end{pmatrix}
+(0,1)(I-e^{-Tt})T^{-1}\begin{pmatrix}a \\ 0\end{pmatrix}\right]\\
&= (0,1)e^{-Tt}\begin{pmatrix}r(\infty) \\ s(\infty)\end{pmatrix}
+m(\infty)(0,1)(I-e^{-Tt})T^{-1}\begin{pmatrix}a \\ 0\end{pmatrix},
\end{split}
\end{equation*}
where we have used the fact that $r(t) = \Enum x(t)\alpha(t)$ and $s(t) = \Enum x(t)^2$ and the fact that the system has reached the steady state. Recall that the autocorrelation function is defined by $C(t) = \Enum x(0)x(t)-\Enum x(0)\Enum x(t)$. Then we finally obtain the explicit expression of the autocorrelation function, which is given by
\begin{equation}\label{correlation}
C(t) = (0,1)e^{-Tt}\begin{pmatrix}r(\infty) \\ s(\infty)\end{pmatrix}
+m(\infty)(0,1)(I-e^{-Tt})T^{-1}\begin{pmatrix}a \\ 0\end{pmatrix}-m(\infty)^2.
\end{equation}
Taking $t = 0$ in the above equation and applying Eqs. \eqref{ssm} and \eqref{sss} yield
\begin{equation}\label{origin}
C(0) = s(\infty)-m(\infty)^2 = \frac{\rho^2da(db+\rho u)}{[d(a+b)+\rho u]^2[d(d+a+b)+\rho u]}.
\end{equation}
Moreover, taking derivatives on both sides of Eq. \eqref{correlation} shows that
\begin{equation*}
C'(t) = -(0,1)e^{-Tt}T\begin{pmatrix}r(\infty) \\ s(\infty)\end{pmatrix}
+m(\infty)(0,1)e^{-Tt}\begin{pmatrix}a \\ 0\end{pmatrix}.
\end{equation*}
Taking $t = 0$ in the above equation yields $C'(0) = 0$, which shows that the autocorrelation function is locally ``flat" around the origin. Here we have used the fact that
\begin{equation*}
T\begin{pmatrix}r(\infty) \\ s(\infty)\end{pmatrix} = 0,
\end{equation*}
which follows easily from Eq. \eqref{sss}.

Without loss of generality, we assume that the two eigenvalues of $T$ are distinct, i.e. $\lambda_1\neq \lambda_2$. In fact, any matrix can be approximated by such matrices to any degree of accuracy. It then follows from Eq. \eqref{correlation} that
\begin{equation*}
C(t) = c_1e^{-\lambda_1t}+c_2e^{-\lambda_2t},
\end{equation*}
where $c_1$ and $c_2$ are two constants. Since $C'(0) = 0$, we have $\lambda_1c_1+\lambda_2c_2 = 0$. Hence the autocorrelation function can be computed explicitly as
\begin{equation}\label{auto}
C(t) = K\left[\frac{\lambda_1e^{-\lambda_2t}-\lambda_2e^{-\lambda_1t}}{\lambda_1-\lambda_2}\right],
\end{equation}
where
\begin{equation}\label{Kexp}
K = C(0) = \frac{\rho^2da(db+\rho u)}{[d(a+b)+\rho u]^2[d(d+a+b)+\rho u]}
\end{equation}
is a constant that can be determined by Eq. \eqref{origin}. Taking the Fourier transform gives the following analytical expression of the power spectrum:
\begin{equation}\label{power}
G(\omega) = \frac{2K\lambda_1\lambda_2(\lambda_1+\lambda_2)}
{(4\pi^2\omega^2+\lambda_1^2)(4\pi^2\omega^2+\lambda_2^2)}.
\end{equation}
It is easy to check that if $\lambda_1,\lambda_2>0$, then both $C(t)$ and $G(\omega)$ are monotonically decreasing.

If the gene is unregulated, then the two eigenvalues of $T$ are given by $\lambda_1 = a+b$ and $\lambda_2 = d$. In this case, the autocorrelation function and power spectrum can be written more explicitly as
\begin{equation}\label{nofeedback}
\begin{split}
&C(t) = \frac{\rho^2ab}{d(a+b)^2(d+a+b)}\left[\frac{de^{-(a+b)t}-(a+b)e^{-dt}}{d-a-b}\right],\\
&G(\omega) = \frac{2\rho^2ab}{(a+b)(4\pi^2\omega^2+d^2)[4\pi^2\omega^2+(a+b)^2]}.
\end{split}
\end{equation}

\subsection{Exact solution of the steady-state protein distribution}
By solving Eq. \eqref{forwardode}, we can also obtain the exact steady-state probability density of the protein concentration, which is a beta distribution (see Appendix A for the proof):
\begin{equation}\label{distribution}
p_{ss}(x) = \frac{\Gamma(\beta)w^{1-\beta}}{\Gamma(\alpha)\Gamma(\beta-\alpha)}
x^{\alpha-1}(w-x)^{\beta-\alpha-1},\;\;\;x<w,
\end{equation}
where $w = \rho/d$ is the maximum protein concentration, and $\alpha$ and $\beta$ are two constants given by
\begin{equation*}
\alpha = \frac{a}{d},\;\;\;\beta = \frac{a+b}{d}+\frac{\rho u}{d^2}.
\end{equation*}
Note that when $u = 0$, our model reduces to the hybrid ODE model for unregulated genes and it has long been proved that the protein concentration has a beta distribution in this special case \cite{karmakar2004graded, raj2006stochastic, dattani2017stochastic}. Our results generalize this result to regulated genes ($u\neq 0$).

In fact, the steady-state protein distribution can be either unimodal or bimodal. From \eqref{distribution}, bimodality occurs if and only if $a<d$ and
\begin{equation}\label{bistable}
u < \frac{d(d-b)}{\rho}.
\end{equation}
In particular, for unregulated genes ($u = 0$), the steady-state distribution is bimodal if and only if both the gene switching rates are smaller than the protein degradation rate, i.e. $a,b<d$. In fact, a similar result has been proved for the discrete CME model of stochastic gene expression \cite{jiao2015distribution}. When the gene inactivation rate is greater than or equal to the protein degradation rate ($b\geq d$), the right-hand side of Eq. \eqref{bistable} is non-positive and thus bimodality can only occur in positive feedback networks ($u<0$). This is consistent with Thomas' first conjecture. Note that the first conjecture may not be true when $b<d$ --- it is well known \cite{grima2012steady, kumar2014exact} that a negative feedback network can also exhibit bimodality when gene switching is relatively slow ($a,b<d$) and when negative feedback is not too strong ($u < d(d-b)/\rho$).

\section{Noise-induced oscillations}

\subsection{Triphasic stochastic bifurcation in negative feedback loops}
We next examine the properties of the autocorrelation function and power spectrum. We first consider the case where the gene is unregulated or positively regulated, i.e. $u\leq 0$. In this case, $\lambda_1$ and $\lambda_2$ are positive real numbers, and hence $C(t)$ and $G(\omega)$ are both monotonically decreasing. This shows that sustained oscillations fail to be observed if there is no negative feedback loops involved in the network. This is consistent with Thomas' second conjecture. From Eq. \eqref{auto}, it is clear that the autocorrelation function is the weighted sum of two exponential functions. This agrees with a recent experiment on nuclear localization of Crz1 protein in response to extracellular calcium \cite{cai2008frequency}, where the authors found that at calcium concentrations greater than 100 mM, the autocorrelation function of localization trajectories is well fitted by a sum of two exponentials. In addition, Eq. \eqref{auto} also shows that the weights of the two exponentials have opposite signs, which suggests that the system is in a nonequilibrium steady state. This is because if a Markovian system is in equilibrium, then the autocorrelation function must be a weight sum of exponential functions with non-positive coefficients \cite{qian2003fundamental, jia2015second}.

For a negative feedback loop $(u>0)$, the discriminant $\Delta = (d-a-b)^2-4\rho u$ may become negative. In particular, the negative feedback strength $u$ has a critical value
\begin{equation*}
u_s = \frac{(d-a-b)^2}{4\rho}.
\end{equation*}
If $u\leq u_s$, then $\Delta\geq 0$ and thus $\lambda_1$ and $\lambda_2$ are positive real numbers. In this case, both $C(t)$ and $G(\omega)$ are monotonically decreasing. If $u>u_s$, then $\Delta<0$ and thus $\lambda_1$ and $\lambda_2$ become conjugated complex numbers. For convenience, we represent them as $\lambda_1 = \alpha-\beta i$ and $\lambda_2 = \alpha+\beta i$, where $\alpha,\beta>0$. In this case, the autocorrelation function given in Eq. \eqref{auto} can be rewritten as
\begin{equation*}
C(t) = Ke^{-\alpha t}\cos(\beta t-\phi),
\end{equation*}
where the constant $K$ is given in Eq. \eqref{Kexp} and the phase $\phi$ satisfies $\tan\phi = \alpha/\beta$. Clearly, $C(t)$ displays damped oscillations and thus must be non-monotonic. Moreover, the power spectrum given in Eq. \eqref{power} can be rewritten as
\begin{equation}\label{negativespectrum}
G(\omega) = \frac{2K(\alpha\cos\phi+\beta\sin\phi)(\alpha^2+\beta^2)}
{[\alpha^2+(2\pi\omega-\beta)^2][\alpha^2+(2\pi\omega+\beta)^2]}.
\end{equation}
It is easy to show that $G(\omega)$ is non-monotonic if and only if (the absolute value of) the imaginary part of $\lambda_1$ and $\lambda_2$ exceeds the real part, i.e. $\beta>\alpha$. Straightforward computations show that
\begin{equation*}
\beta^2-\alpha^2 = \frac{1}{2}[2\rho u-(a+b)^2-d^2].
\end{equation*}
Therefore, there is another critical value
\begin{equation}\label{critical}
u_c = \frac{(a+b)^2+d^2}{2\rho} > u_s = \frac{(d-a-b)^2}{4\rho}
\end{equation}
for the negative feedback strength $u$. If $u_s<u\leq u_c$, then $\beta\leq\alpha$ and $G(\omega)$ must be monotonically decreasing. In this case, while the autocorrelation function displays damped oscillations, the power spectrum is still monotonic and thus oscillations cannot be observed. If $u>u_c$, then $\beta>\alpha$ and thus $G(\omega)$ becomes non-monotonic. In particular, a negative feedback network with inversely correlated gene switching rates can exhibit robust oscillations when the feedback strength is sufficiently large. In this case, $G(\omega)$ attains its maximum at
\begin{equation*}
\omega = \frac{\sqrt{\beta^2-\alpha^2}}{2\pi} = \frac{\sqrt{\rho(u-u_c)}}{2\pi},
\end{equation*}
which can be understood as the dominant frequency of oscillations. From Eq. \eqref{critical}, it is clear that both a large protein degradation rate $d$ and a large total gene switching rate $a+b$ will result in a large critical value $u_c$. This implies that sustained oscillations is almost impossible to occur when the protein is very unstable or when gene switching is very fast. On the contrary, oscillations tend to take place when the protein is relatively stable and when gene switching is relatively slow.
\begin{figure}[!thb]
\centerline{\includegraphics[width=1.0\textwidth]{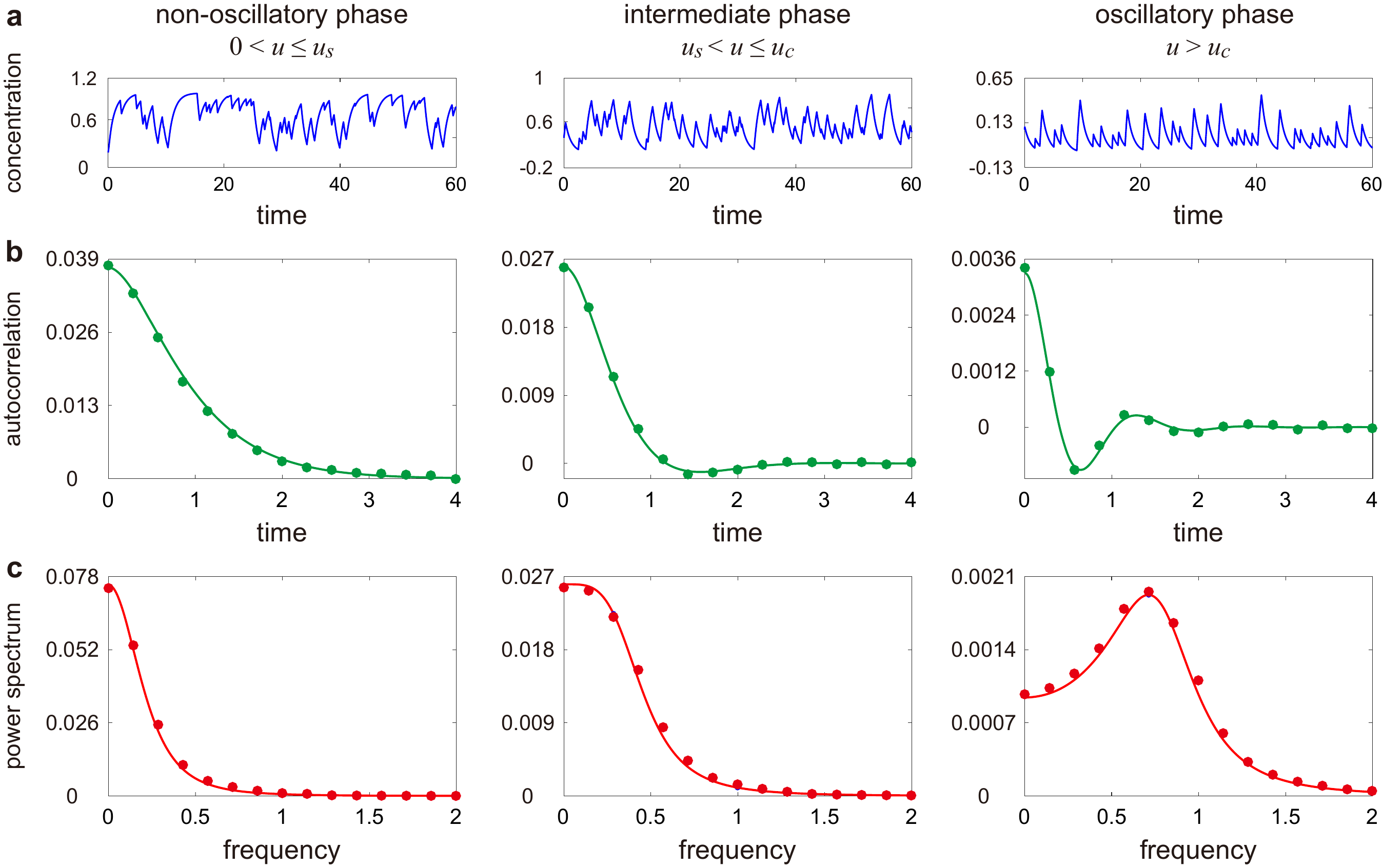}}
\caption{\textbf{Stochastic bifurcations of oscillations in negative feedback networks.} The negative feedback strength $u$ has two critical values $u_s$ and $u_c$, which separate the parameter region into three phases: the non-oscillatory phase of $0<u\leq u_s$, the transitional phase of $u_s<u\leq u_c$, and the oscillatory phase of $u>u_c$.
\textbf{(a)} Typical trajectories of the protein concentration in the three phases.
\textbf{(b)} Autocorrelation functions of concentration fluctuations in the three phases.
\textbf{(c)} Power spectra of concentration fluctuations in the three phases. In (b),(c), the curve shows the analytical solution given in Eqs. \eqref{auto} and \eqref{power}, and the dots show the simulated results. The model parameters in (a)-(c) are chosen as $a = 3, b = 0, \rho = d = 1$. The negative feedback strength is chosen to be $u = u_s$ in the non-oscillatory phase, $u = u_c$ in the transitional phase, and $u = 5u_c$ in the oscillatory phase.} \label{bifurcation}
\end{figure}

We emphasize that here oscillations are noise-induced in the sense that they are found in parameter regions where the deterministic theory predicts stable fixed points. In other words, stochasticity can alter the qualitative behavior predicted by the deterministic rate equation. Such noise-induce oscillations have been extensively studied in the literature using various approximation methods such as the LNA \cite{mckane2007amplified, dauxois2009enhanced, realpe2012demographic, thomas2012slow, toner2013effects}, corrected LNA \cite{thomas2013signatures}, large delay approximation \cite{bratsun2005delay}, and Pad\'{e} approximation \cite{gupta2022frequency}. Here we provide an exact theory in a single feedback loop.

We have now provided a complete characterization of the oscillatory behavior of stochastic gene expression in three types of gene networks. If the gene is unregulated or positively regulated, both $C(t)$ and $G(\omega)$ are monotonic, and thus no oscillations could be observed. However, a stochastic bifurcation will occur if the gene is negatively regulated. To gain an intuitive picture of this, we simulate the trajectories of the hybrid model under three sets of biologically relevant parameters and then use them to compute $C(t)$ and $G(\omega)$ (Fig. \ref{bifurcation}(a)-(c)). The numerical power spectrum is computed by means of the Wiener-Khinchin theorem, which states that $G(\omega) = \lim_{T\rightarrow\infty}\langle|\hat{n}_T(\omega)|^2\rangle/T$, where $\hat{n}_T(\omega) = \int_0^Tn(t)e^{-2\pi i\omega t}dt$ is the truncated Fourier transform of a single stochastic trajectory over the interval $[0,T]$ and the angled brackets denote the ensemble average over trajectories. In fact, the two critical values of the negative feedback strength, $u_s$ and $u_c$, separate the parameter region into three phases. In the non-oscillatory phase of $u\leq u_s$, both $C(t)$ and $G(\omega)$ are monotonic. In the transitional phase of $u_s<u\leq u_c$, the former becomes non-monotonic while the latter is still monotonic. In the oscillatory phase of $u>u_c$, both quantities become non-monotonic. Note that when the protein is very stable ($d\ll 1$), the critical value $u_c = (a+b)^2/2\rho$ separating the transitional and oscillatory phases is twice the critical value $u_s = (a+b)^2/4\rho$ separating the transitional and non-oscillatory phases.

Let $G_{\textrm{peak}}$ denote the peak value of the power spectrum. To evaluate the performance of stochastic oscillations, we define a quantity referred to as \emph{oscillatory efficiency}:
\begin{equation*}
\eta = \frac{G_{\textrm{peak}}-G(0)}{G_{\textrm{peak}}} = 1-\frac{G(0)}{G_{\textrm{peak}}},
\end{equation*}
which is a number between 0 and 1. Clearly, a large value of $\eta$ implies that the spectrum at the dominant frequency is much larger than the spectrum at zero frequency, which means that oscillations are strong. If $G(\omega)$ is monotonically decreasing, then $G_{\textrm{peak}} = G(0)$ and thus $\eta$ vanishes. Hence the efficiency $\eta$ serves as an effective indicator that describes the robustness of oscillations. When $u>u_c$, the efficiency can be computed explicitly as
\begin{equation*}
\eta = 1-\left[\frac{2\alpha\beta}{\alpha^2+\beta^2}\right]^2
= \left[\frac{2\rho(u-u_c)}{(d+a+b)^2+2\rho(u-u_c)}\right]^2.
\end{equation*}
As the negative feedback strength $u$ increases, the efficiency $\eta$ becomes larger and thus oscillations become more apparent (Fig. \ref{bifurcation}(a)).

Experimentally, the single-cell tracks of gene expression levels constantly fluctuate around the mean. For such time-lapse data, it is sometimes difficult to distinguish whether the underlying dynamics belongs to stochastic bursts \cite{suter2011mammalian} or stochastic oscillations. This issue has been pointed out in \cite{lin2018efficient}, which says ``This is difficult to answer because the distinction between stochastic cycling and oscillation is ill-defined." This has also been noticed in a recent experiment on nuclear localization of Crz1 protein \cite{cai2008frequency}, where the authors found that at very high calcium levels, the Crz1 nuclear localization bursts look similar to sustained oscillations. In fact, our theory gives a clear distinction between stochastic bursts ($u\leq u_c$) and stochastic oscillations ($u>u_c$). Typically, stochastic bursts do not have an intrinsic frequency and thus give rise to a monotonic power spectrum (while it may have an oscillatory autocorrelation function), whereas stochastic oscillations have an intrinsic period and are featured by a non-monotonic power spectrum \cite{bratsun2005delay, mckane2007amplified, realpe2012demographic, thomas2012slow, toner2013effects, thomas2013signatures, thomas2014phenotypic, jia2021frequency, jia2022concentration, gupta2022frequency}.

\begin{figure}[!thb]
\centerline{\includegraphics[width=0.8\textwidth]{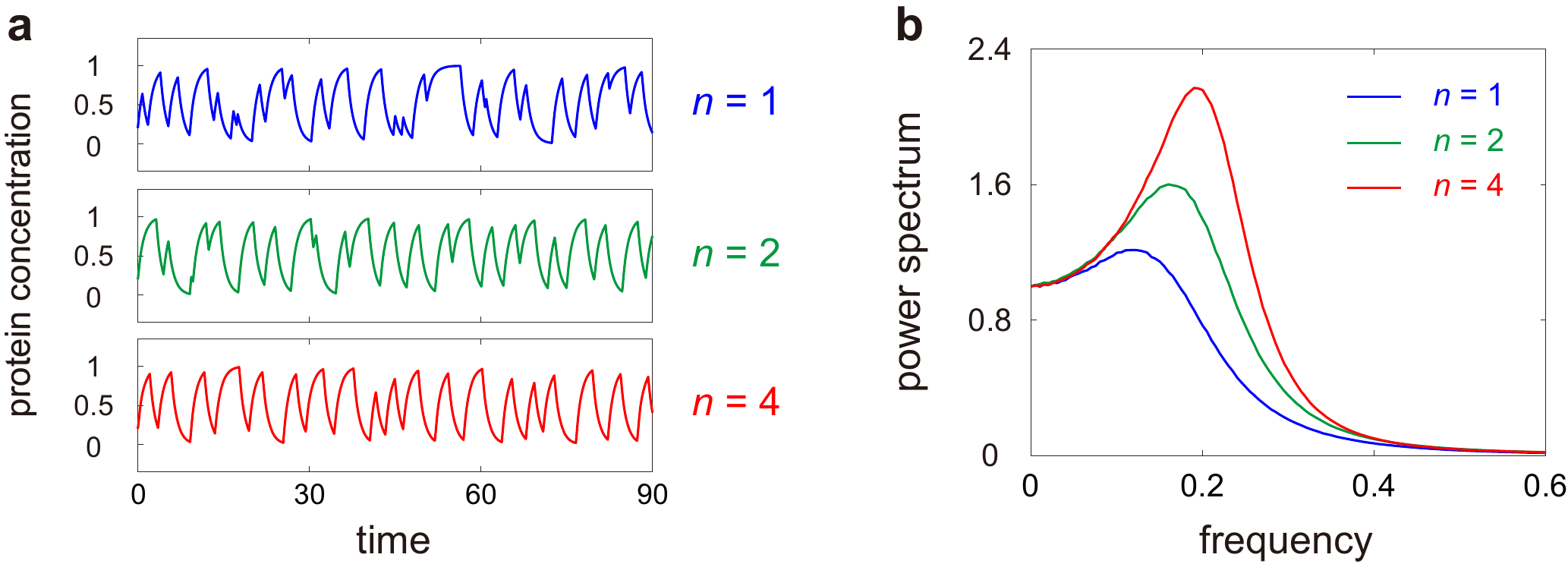}}
\caption{\textbf{Oscillations when gene switching rates depend nonlinearly on the protein concentration.} Here the gene switching rates are chosen as Hill functions $a(x) = a/[1+(ux)^n]$ and $b(x) = a/[1+(u(w-x))^n]$, where $w = \rho/d$ is the maximum protein concentration and $u$ is the strength of negative feedback. Note that when the cooperativity $n$ is large, $a(x)$ transitions sharply from $a$ to $0$ around $x = 1/u$, while $b(x)$ transitions sharply from $0$ to $a$ around $x = w-1/u$ (see Fig. \ref{hysteresis}(b) for an illustration).
\textbf{(a)} Typical trajectories of the protein concentration when $n = 1,2,4$.
\textbf{(b)} Simulated power spectra of concentration fluctuations when $n = 1,2,4$. Here the power spectra are normalized so that $G(0) = 1$. The model parameters in (a),(b) are chosen as $a = \rho = d = 1, u = 5$.} \label{hill}
\end{figure}

Thus far, our analytical solution is derived under the assumption that the two gene switching rates linearly depend on the protein concentration and are inversely correlated (Eq. \eqref{functional}). However, our main results are actually insensitive to the specific functional forms of $a(x)$ and $b(x)$ except their monotonicity. More generally, we may assume that $a(x) = a-vx$ and $b(x) = b+ux$, where $u$ and $v$ are two constants jointly characterizing the feedback strength. In this case, it is almost impossible to obtain the analytic solutions of $C(t)$ and $G(\omega)$. However, according to simulations, a negative feedback network also undergoes a triphasic stochastic bifurcation as $u$ and $v$ increases while keeping $u/v$ as constant. Furthermore, robust oscillations can also been observed when $a(x)$ and $b(x)$ are chosen to be nonlinear Michaelis-Menten or Hill functions with opposite monotonicity, and oscillations become stronger as the cooperativity in the Hill function increases (Fig. \ref{hill}). Unfortunately, we find that sustained oscillations can never be observed if $a(x) = a$ is a constant and $b(x) = b+ux$, an assumption imposed on many models of autorepressive genes \cite{hornos2005self, grima2012steady, kumar2014exact, jia2020small}. This indicates that an oversimple regulation mechanism may not be sufficient to yield robust oscillations.

\subsection{Deeper understanding of Thomas' conjectures}
\begin{figure}[!htb]
\centerline{\includegraphics[width=1.0\textwidth]{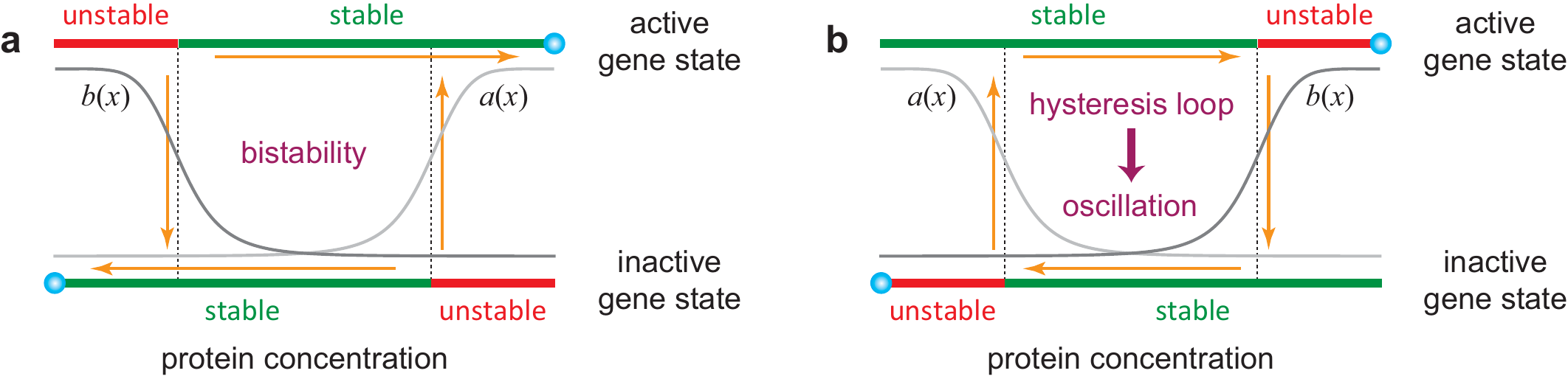}}
\caption{\textbf{Schematics of the dynamic evolution of the hybrid model in positive and negative feedback loops.}
\textbf{(a)} Positive feedback loops.
\textbf{(b)} Negative feedback loops.
The upper and lower layers represent the active and inactive gene states. The light and dark grey curves show the graphs of the gene switching rates $a(x)$ and $b(x)$, respectively. The blue balls show the locations of the stable attractors for the two gene states. The stable and unstable regions for the two gene states are depicted as green and red bars, respectively. The yellow arrows indicate the evolution direction of the Markovian dynamics.}\label{hysteresis}
\end{figure}

Based on the stochastic hybrid model, Thomas' two conjectures can be understood intuitively using the schematic diagrams depicted in Fig. \ref{hysteresis}. Since the gene switching rates $a(x)$ and $b(x)$ have opposite monotonicity in positive and negative feedback networks, the stable and unstable regions for the active and inactive gene states are also different in the two types of networks, each forming a stochastic hysteresis loop. Here stable and unstable regions are defined as regions with slow and fast gene switching rates, respectively. For positive feedback loops, the stable attractors for the two gene states lie in the stable regions, forming bimodality (Fig. \ref{hysteresis}(a)). For negative feedback loops, the stable attractors lie in the unstable regions. Before the protein level could approach a stable attractor, it has already entered the unstable region and thus will transition to the other gene state. With time, the protein level will fluctuate around the hysteresis loop, forming sustained oscillations (Fig. \ref{hysteresis}(b)). This understanding provides a theoretical complement to previous studies on relaxation oscillators, where similar hysteretic switch was found \cite{tsai2008robust}. However, the difference is that the hysteresis in relaxation oscillators comes from an additional positive feedback loop \cite{tsai2008robust}, while in our hybrid model (Fig. \ref{hysteresis}(b)), the network only has a negative feedback loop and the hysteresis is induced by gene state switching. In addition, relaxation oscillations are characterized by two alternating processes on different time scales: a long relaxation period during which the system approaches an equilibrium point, alternating with a short impulsive period in which the equilibrium point shifts \cite{ginoux2012van}. However, in our hybrid model, timescale separation is not necessary; on the contrary, very fast gene switching or protein degradation will destroy oscillations.

\section{Generalization to systems with bursty protein synthesis}
In single cells, protein is often produced in a bursty manner \cite{cai2006stochastic}. Both non-bursty and bursty gene expression are commonly observed in naturally occurring systems \cite{zenklusen2008single}. When protein synthesis is bursty, the evolution of the protein concentration $x$ can be modeled by a hybrid stochastic differential equation (SDE) \cite{jia2017emergent, jia2019single}:
\begin{equation*}
\xymatrix@R = 0.5cm{
\dot{x} = \dot{\xi}_t-dx \ar@<0.2ex>@^{->}[d]^{b(x)} & (\textrm{active state}),\\
\dot{x} = -dx \ar@<0.2ex>@^{->}[u]^{a(x)} & (\textrm{inactive state}).}
\end{equation*}
where $\xi_t$ is a compound Poisson process capturing bursty production of protein (Fig. \ref{network}(c)). The jump rate of $\xi_t$ corresponds to the burst frequency and the jump distribution of $\xi_t$ corresponds to the burst size distribution. Note that when the gene is always active and the burst size has an exponential distribution, this model reduces to the classical bursty model of gene expression proposed in \cite{friedman2006linking} (also see \cite{mackey2013dynamic, bokes2015protein, jkedrak2016time, chen2020limit}). Recently, it has been shown \cite{jia2017emergent, jia2019single} that the hybrid ODE and SDE models can be viewed as different approximations of the discrete CME model of stochastic gene expression when protein numbers are relatively large. The former works better in the regime of large burst frequencies, while the latter works better in the regime of large burst sizes.

The evolution of the hybrid SDE model is governed by the Kolmogorov forward equation
\begin{equation}\label{forwardsde}\left\{
\begin{split}
\partial_tp_0 &= d\partial_x(xp_0)+b(x)p_1-a(x)p_0,\\
\partial_tp_1 &= d\partial_x(xp_1)+s\mu*p_1+a(x)p_0-[b(x)+s]p_1,
\end{split}\right.
\end{equation}
where $s$ is the burst frequency, $\mu(dx)$ is the burst size distribution, and $\mu*p_1(x) = \int_0^xp_1(x-y)\mu(dy)$ denotes the convolution of $\mu$ and $p_1$. In previous studies \cite{friedman2006linking}, the burst size distribution is often chosen to be the exponential distribution $\mu(dx) = \lambda e^{-\lambda x}dx$, which is supported by experiments \cite{cai2006stochastic}. Similarly to the non-bursty case, both the autocorrelation function $C(t)$ and power spectrum $G(\omega)$ can be calculated explicitly when protein synthesis is bursty. If the gene is unregulated, $C(t)$ and $G(\omega)$ are given by (see Appendix B for the proof)
\begin{equation}\label{nofeedbackmod}
C(t) = \widehat C(t)+\widetilde C(t),\;\;\;G(\omega) = \widehat G(\omega)+\widetilde G(\omega),
\end{equation}
where $\widehat C(t)$ and $\widehat G(\omega)$ are the functions given in Eq. \eqref{nofeedback} with $\rho = sB$, and
\begin{equation*}
\widetilde C(t) = \frac{as\sigma}{2d(a+b)}e^{-dt},\;\;\;
\widetilde G(\omega) = \frac{as\sigma}{(a+b)(4\pi^2\omega^2+d^2)}
\end{equation*}
with $\sigma = \int_0^\infty x^2\mu(dx)$ being the second moment of the burst size distribution. In this case, both $C(t)$ and $G(\omega)$ are the sums of two monotonically decreasing functions. This leads to the same conclusion that oscillations cannot be observed if the gene is unregulated, even if protein bursting is taken into account.

In the presence of feedback loops, $C(t)$ and $G(\omega)$ can also be calculated analytically as
\begin{equation}\label{positivemod}
C(t) = c_1e^{-\lambda_1t}+c_2e^{-\lambda_2t},\;\;\;
G(\omega) = \frac{2c_1\lambda_1}{4\pi^2\omega^2+\lambda_1^2}
+\frac{2c_2\lambda_2}{4\pi^2\omega^2+\lambda_2^2},
\end{equation}
where $c_1$ and $c_2$ can be determined by solving the following set of linear equations:
\begin{equation}\label{eq1}
\begin{split}
&c_1+c_2 = \frac{2\rho^2da(db+\rho u)+das\sigma(d+a+b)[d(a+b)+\rho u]}{2[d(a+b)+\rho u]^2[d(d+a+b)+\rho u]},\\
&\lambda_1c_1+\lambda_2c_2 = \frac{das\sigma}{2[d(a+b)+\rho u]},
\end{split}
\end{equation}
with $\rho = sB$ being the effective synthesis rate of protein. In this case, the autocorrelation function (power spectrum) is still the weighted sum of two exponential (Lorentzian) functions, but the weights are much more complicated than those in the hybrid ODE model. Straightforward computations show that the expressions of $C(t)$ and $G(\omega)$ in the bursty case (Eq. \eqref{positivemod}) reduce to those in the non-bursty case (Eqs. \eqref{auto} and \eqref{power}) when the second moment of the burst size distribution vanishes, i.e. $\sigma = 0$.
\begin{figure}[!htb]
\centerline{\includegraphics[width=0.8\textwidth]{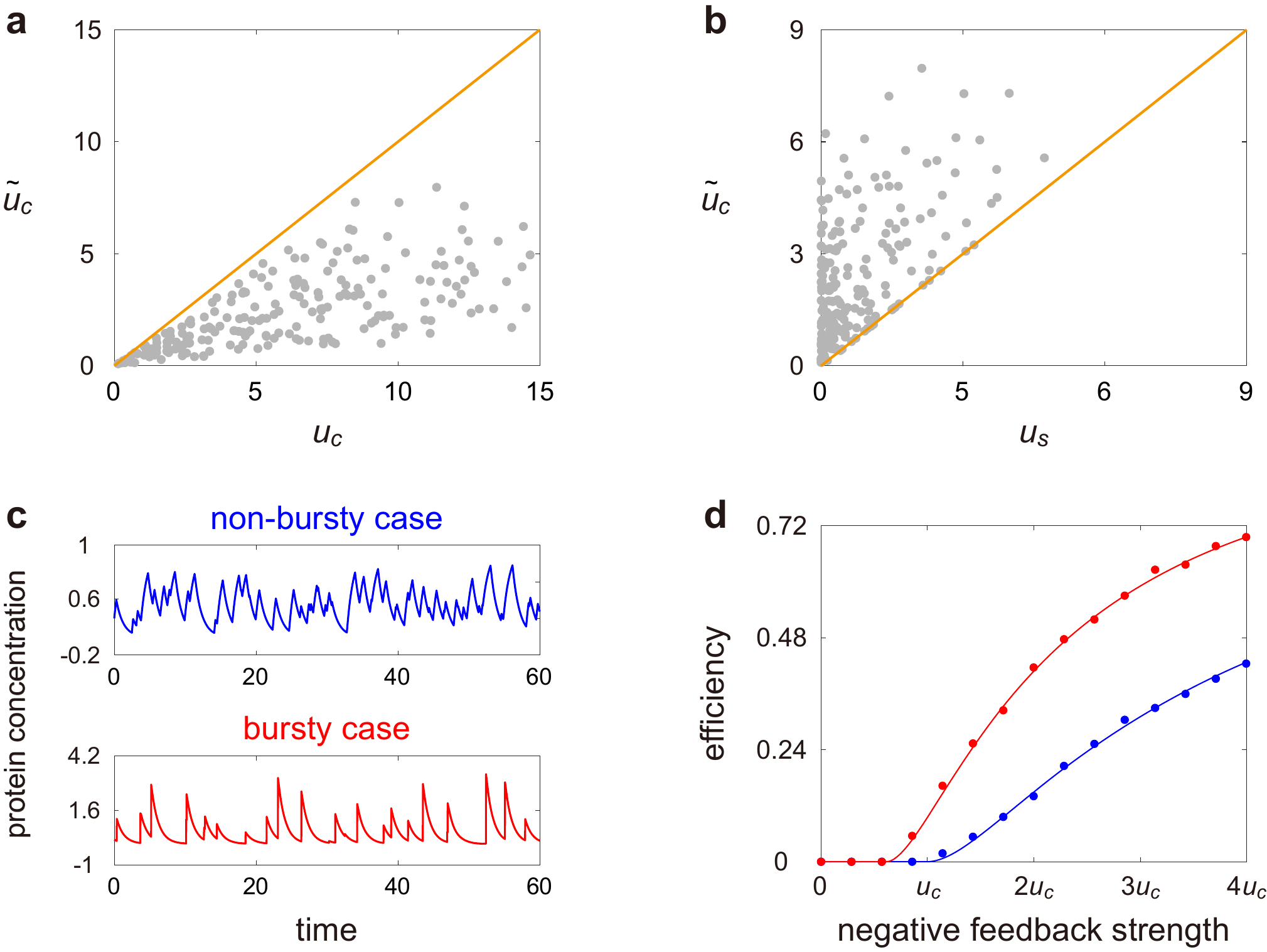}}
\caption{\textbf{Comparison of stochastic oscillations in the non-bursty and bursty cases.} Stochastic bifurcations of oscillations can be observed in both the non-bursty and bursty models. In the bursty case, the burst size is chosen to be exponentially distributed.
\textbf{(a)} Scatter plot of $\tilde{u}_c$ versus $u_c$, where $\tilde{u}_c$ is critical value of oscillations in the bursty model.
\textbf{(b)} Scatter plot of $\tilde{u}_c$ versus $u_s$. In (a),(b), we randomly sample $200$ sets of model parameters. Specifically, the model parameters are randomly chosen as $a\sim U[0,10], b\sim U[0,1], s\sim U[0,10], sB\sim U[1,10], d\sim U[0,10]$, where $U[x,y]$ denotes the uniform distribution over the interval $[x,y]$.
\textbf{(c)} Typical trajectories for the non-bursty and bursty models when $u = u_c$.
\textbf{(d)} Oscillatory efficiencies for the non-bursty (blue) and bursty (red) models as a function of the negative feedback strength $u$. In (c),(d), the curves show theoretical results and the dots show simulated ones. The model parameters are chosen as $a = 3, b = 0, d = 1$ in both the non-bursty and bursty cases, $\rho = 1$ in the non-bursty case, and $s = B = 1$ in the bursty case.}\label{comparison}
\end{figure}

To compare the oscillatory behaviors of the hybrid ODE and SDE models, we choose model parameters so that they share the same mean field dynamics. According to simulations, we find that stochastic bifurcations take place in both models. However, for the hybrid SDE model, it is almost impossible to compute the explicit expression of the critical value. Interestingly, we find that the critical value of oscillations $\tilde u_c$ (the critical value separating monotonic and non-monotonic power spectra) in the bursty case is always smaller than its counterpart $u_c$ in the non-bursty case, suggesting that the hybrid SDE model requires a smaller negative feedback strength $u$ to generate robust oscillations than the hybrid ODE model. To see this, we randomly select $200$ sets of model parameters and compute the values of $u_s$, $u_c$, and $\tilde{u}_c$ under each set of parameters (the explicit expression of the power spectrum given in Eq. \eqref{positivemod} allows us to compute the value of $\tilde{u}_c$ without simulating the trajectories). Fig. \ref{comparison}(a),(b) illustrate the scatter plots of $\tilde{u}_c$ versus $u_c$ and $u_s$. It is clear that for all parameter sets, $\tilde u_c$ is always smaller than $u_c$ and is always larger than $u_s$.

Fig. \ref{comparison}(c) illustrates the typical trajectories for the two models when $u = u_c$. It is clear that oscillations fail to be observed at the critical value $u_c$ for the hybrid ODE model, while the hybrid SDE model displays much more apparent oscillations. To reinforce this result, we depict the oscillatory efficiencies of the two models in Fig. \ref{comparison}(d) as a function of the negative feedback strength $u$. Under the parameters chosen, the critical value in the bursty case is only $0.6u_c$, and for a fixed feedback strength $u$, the hybrid SDE model possesses a much higher efficiency. All these results reveal that burstiness in protein production is expected to boost the occurrence and broaden the region of stochastic oscillations, while it gives rise to larger gene expression noise \cite{paulsson2005models}. A reasonable explanation for this counterintuitive result is that a single burst will drive the protein abundance to jump from a low to a much higher level. This helps the system transition more easily from the stable region for the active gene state to the unstable region, as illustrated in Fig. \ref{hysteresis}(b). Once a burst occurs, the gene will switch rapidly from the active to the inactive state and then the protein abundance will decay to a lower level again to finish a cycle. Hence protein bursting plays an important role in prolonging the cycle times and enhancing the robustness of oscillations (Fig. \ref{comparison}(c)). We emphasize that while bursting promotes oscillations in our model, it may destroy oscillations in other systems --- it has been shown that burstiness may either promote or destroy oscillations in more complex networks involving autocatalysis, trimerization, and feedback loops \cite{toner2013effects}.

\section{Conclusions and discussion}
Stochastic oscillations in single cells are usually characterized by a non-monotonic power spectrum with an oscillatory autocorrelation function. In recent years, numerous efforts have been made to analytically compute the power spectrum in stochastic biochemical networks \cite{bratsun2005delay, mckane2007amplified, realpe2012demographic, thomas2012slow, toner2013effects, thomas2013signatures, thomas2014phenotypic, jia2021frequency, jia2022concentration, gupta2022frequency}. The difficulty in obtaining the analytical power spectrum arises from the fact that the moment equations are not closed in the presence of feedback loops. To solve this, various approximation methods have been proposed, including the LNA \cite{mckane2007amplified, dauxois2009enhanced, realpe2012demographic, thomas2012slow, toner2013effects}, corrected LNA \cite{thomas2013signatures}, large delay approximation \cite{bratsun2005delay}, and Pad\'{e} approximation \cite{gupta2022frequency}. However, even for a simple autoregulatory loop, there is no exact solution for the power spectrum.

In this paper, we developed an exact theory of stochastic oscillations in a minimal stochastic gene expression model including gene state switching, protein synthesis and degradation, as well as a genetic feedback loop. The positive or negative feedback is modelled as linear dependence of the gene activation and inactivation rates on the protein concentration with opposite monotonicity. The evolution of the system is modelled as a hybrid ODE, also called a piecewise-deterministic Markov process, where the gene state is modeled as a discrete variable and the protein concentration is modeled as a continuous variable. In our model, the zeroth moment in the active gene state and the first moment of concentration fluctuations satisfy a closed moment equation. Moreover, the first moment in the active gene state and the second moment also satisfy a closed moment equation. Using these two closed moment equations and the technique of conditional expectation, we derived the exact analytical solutions of the autocorrelation function and power spectrum in the presence of feedback regulation. Moreover, we also computed the steady-state protein distribution in closed form and found that the protein concentration has a beta distribution. We emphasize that while the protein abundance is modeled continuously in the present work, the exact autocorrelation and spectrum can be computed using similar methods if both the gene state and protein abundance are modelled discretely.

Note that hybrid ODE models have been used previously to describe stochastic oscillations in biological and ecological systems \cite{realpe2012demographic, lin2018efficient}. Ref. \cite{lin2018efficient} explains the origins of stochastic oscillations, especially in the regime of slow gene switching; however, it does not discuss the power spectrum. Ref. \cite{realpe2012demographic} computed the approximate power spectrum using the LNA. Note that in \cite{realpe2012demographic}, the discrete variable in the hybrid model represents the number of plants and hence its dynamics can be approximated by the LNA when the total area of the system is large \cite{realpe2012demographic}. However in our hybrid model, the discrete variable, i.e. the gene state, can only take two values ($i = 0,1$) and hence the conventional LNA is not valid.

Our analytical results reinforce previous results that robust oscillations fail be observed when the gene is unregulated or positively regulated since the eigenvalues of the characteristic matrix are both real. We observed a triphasic stochastic bifurcation in negative feedback networks as the feedback strength increases: the autocorrelation function first changes from a monotonic to a non-monotonic shape at a small critical value due to the emergence of complex eigenvalues, and then the power spectrum changes from a monotonic to a non-monotonic shape at a larger critical value due to the competition between the real and imaginary parts of the eigenvalues. The two critical values separate the parameter region into a non-oscillatory phase with monotonic autocorrelation and spectrum, a transitional phase with non-monotonic autocorrelation and monotonic spectrum, and an oscillatory phase with non-monotonic autocorrelation and spectrum. Our analytical results indicate that very fast gene switching or protein degradation will destroy oscillations. According to simulations, we further showed that similar stochastic bifurcations can also be observed when the gene switching rates depend nonlinearly on the protein concentration, and oscillations become stronger as nonlinearity increases.

Our hybrid model provides a deeper understanding of Thomas' two conjectures. Since the gene switching rates have opposite dependence on the protein concentration in positive and negative feedback networks, the two gene states also have different stable and unstable regions in the two types of networks, each forming a stochastic hysteresis loop. For positive feedback loops, the attractors for the two gene states lie in the stable regions, forming bimodality. For negative feedback loops, the attractors for the two gene states lie in the unstable regions; in this case, the protein level will fluctuate around the hysteresis loop, forming sustained oscillations. The essence of stochastic oscillations in our model is revealed to be a circular motion along the stochastic hysteresis loop. The hysteresis in conventional relaxation oscillators comes from an additional positive feedback loop \cite{tsai2008robust}, while in our hybrid model, the hysteresis is induced by gene state switching.

Finally, we investigated the influence of translational bursting on stochastic oscillations by modelling protein synthesis as a compound Poisson process with the jump rate corresponding to the burst frequency and with the jump distribution corresponding to the burst size distribution. This yields a hybrid SDE model. Similarly to the non-bursty case, we computed the exact autocorrelation function and power spectrum when protein synthesis occurs in a bursty manner. Based on theory and simulations, we showed that the critical value of oscillations in the bursty case is always lower than its counterpart in the non-bursty case. Moreover, when the feedback strength is fixed, the bursty model always has a higher oscillatory efficiency compared to the non-bursty model. These results reveal that even though bursting yields larger gene expression noise, it can broaden the region and enhance the robustness of stochastic oscillations in a simple negative feedback loop.

In summary, previous studies have proposed various approximation methods to compute the power spectrum of concentration fluctuations in stochastic nonlinear biochemical networks. For example, the LNA \cite{mckane2007amplified, dauxois2009enhanced, realpe2012demographic, thomas2012slow, toner2013effects} and corrected LNA \cite{thomas2013signatures} can be used to derive the analytical power spectrum for general biochemical networks in the vicinity of a Hopf bifurcation when the system size is large; the Pad\'{e} approximation \cite{gupta2022frequency} can be used to obtain the semi-analytical power spectrum in terms of rational functions for general biochemical networks. In this paper, we derived the analytical power spectrum in a simple genetic feedback loop. The advantage of our method is that it does not require any approximations and hence is exact. However, the model considered here is too simple compared to those studied in previous papers. We anticipate that the method developed in this paper can be generalized to gene expression models with complex promoter switching \cite{jiao2020regulation, chen2022using, jia2023analytical}, gene expression models with time delay \cite{bratsun2005delay, stricker2008fast}, as well as gene networks with complex feedback regulation \cite{thomas2014phenotypic, gupta2022frequency}.

\section*{Appendices}

\subsection*{A. Derivation of the steady-state protein distribution}
Let $p^{ss}_i(x)$ denote the steady-state probability density of the protein concentration when the gene is in state $i$, and let $p^{ss}(x) = p^{ss}_0(x)+p^{ss}_1(x)$. In steady state, adding the two equations in Eq. \eqref{forwardode} yields
\begin{equation*}
\frac{d}{dx}[dxp^{ss}_0-(\rho-dx)p^{ss}_1] = 0.
\end{equation*}
This shows that
\begin{equation*}
dxp^{ss}_0-(\rho-dx)p^{ss}_1 = K,
\end{equation*}
where $K$ is a constant. Note that $p^{ss}_0$ and $p^{ss}_1$ must decay exponentially as $x\rightarrow\infty$. Taking $x\rightarrow\infty$ in the above equation shows that $K = 0$. Hence we have $(\rho-dx)p^{ss}_1(x) = dxp^{ss}_0(x)$. For convenience, we set
\begin{equation*}
\psi(x) = (\rho-dx)p^{ss}_1(x) = dxp^{ss}_0(x).
\end{equation*}
Moreover, it follows from Eq. \eqref{forwardode} that
\begin{equation*}
\psi'(x) = a(x)p^{ss}_0(x)-b(x)p^{ss}_1(x) = \left[\frac{a(x)}{dx}-\frac{b(x)}{\rho-dx}\right]\psi(x).
\end{equation*}
Integrating the above equation gives
\begin{equation}\label{step1}
\psi(x) = C\exp\left\{\int_1^x \left[\frac{a(y)}{dy}-\frac{b(y)}{\rho-dy}\right]dy\right\},
\end{equation}
where $C$ is a normalization constant. Finally, the steady-state probability density of the protein concentration is given by
\begin{equation}\label{step2}
p^{ss}(x) = p^{ss}_0(x)+p^{ss}_1(x) = \left[\frac{1}{dx}+\frac{1}{\rho-dx}\right]\psi(x).
\end{equation}
Computing the integral in Eq. \eqref{step1} explicitly and inserting the resulting $\psi(x)$ into Eq. \eqref{step2} gives Eq. \eqref{distribution} in the main text.

\subsection*{B. Derivation of the autocorrelation function and power spectra in the bursty case}
The computation of the autocorrelation function and power spectra for the hybrid SDE model follows the same line as that for the hybrid ODE model. Integrating the second identity in Eq. \eqref{forwardsde}, we obtain
\begin{equation*}
\begin{split}
\frac{d}{dt}g(t) &= \int_0^\infty\partial_tp_1dx
= s\int_0^\infty\mu*p_1dx+\int_0^\infty[(a-ux)p_0-(b+s+ux)p_1]dx \\
&= s\int_0^\infty\mu(dy)\int_y^\infty p_1(x-y,t)dx+\int_0^\infty[ap_0-(b+s)p_1-uxp]dx \\
&= sg(t)+a[1-g(t)]-(b+s)g(t)-um(t) \\
&= a-(a+b)g(t)-um(t).
\end{split}
\end{equation*}
Similarly, using Eq. \eqref{forwardsde} and integration by parts, we obtain
\begin{equation*}
\begin{split}
\frac{d}{dt}m(t) &= \int_0^\infty x\partial_tpdx
= \int_0^\infty x[d\partial_x(xp)+s\mu*p_1-sp_1]dx \\
&= s\int_0^\infty\mu(dy)\int_y^\infty xp_1(x-y,t)dx-s\int_0^\infty p_1dx-d\int_0^\infty xpdx \\
&= s\int_0^\infty\mu(dy)\int_0^\infty(x+y)p_1(x,t)dx-s\int_0^\infty p_1dx-d\int_0^\infty xpdx \\
&= \rho g(t)-dm(t),
\end{split}
\end{equation*}
where $\rho = s\int_0^\infty y\mu(dy)$ is the product of the burst frequency and mean burst size. In this case, $g(t)$ and $m(t)$ also satisfy the system of ODEs given in Eq. \eqref{ODE}. Let $m_i(x,t)$ denote the conditional mean of the protein concentration at time $t$ given that the initial gene state is $i$ and the initial protein concentration is $x$. Then $m_i(x,t)$ can be computed explicitly as in Eq. \eqref{conditional}.

To proceed, let
\begin{equation*}
r(t) = \int_0^\infty xp_1(x,t)dx,\;\;\;s(t) = \int_0^\infty x^2p(x,t)dx.
\end{equation*}
Then we have
\begin{equation*}
\begin{split}
\frac{d}{dt}r(t) &= \int_0^\infty x\partial_tp_1dx
= \int_0^\infty x[d\partial_x(xp_1)+s\mu*p_1+(a-ux)p_0-(b+s+ux)p_1]dx \\
&= \int_0^\infty[sx\mu*p_1-dxp_1+axp_0-(b+s)xp_1-ux^2p]dx \\
&= \rho g(t)+am(t)-(d+a+b)r(t)-us(t).
\end{split}
\end{equation*}
Similarly, we have
\begin{equation*}
\begin{split}
\frac{d}{dt}s(t) &= \int_0^\infty x^2\partial_tpdx
= \int_0^\infty x^2[d\partial_x(xp)+s\mu*p_1-sp_1]dx \\
&= s\int_0^\infty\mu(dy)\int_y^\infty x^2p_1(x-y,t)dx-s\int_0^\infty x^2p_1dx-2d\int_0^\infty x^2pdx \\
&= s\int_0^\infty\mu(dy)\int_0^\infty(x+y)^2p_1(x,t)dx-s\int_0^\infty x^2p_1dx-2d\int_0^\infty x^2pdx \\
&= 2\rho\int_0^\infty xp_1(x,t)dx+s\int_0^\infty y^2\mu(dy)\int_0^\infty p_1(x,t)dx
-2d\int_0^\infty x^2pdx \\
&= 2\rho r_1(t)+\gamma g(t)-2ds(t),
\end{split}
\end{equation*}
where $\gamma = s\int_0^\infty y^2\mu(dy)$ is the product of the burst frequency and the second moment of the burst size. This shows that
\begin{equation}\label{ODE2mod}
\frac{d}{dt}\begin{pmatrix}r(t) \\ s(t)\end{pmatrix}
= -R\begin{pmatrix}r(t) \\ s(t)\end{pmatrix}
+\begin{pmatrix}\rho g(t)+am(t) \\ \gamma g(t)\end{pmatrix},
\end{equation}
where $R$ is the matrix given in Eq. \eqref{Rmatrix}. Note that in the busty case, the above equation is different from Eq. \eqref{ODE2} in the non-bursty case. In steady state, it follows from Eq. \eqref{ODE} that
\begin{equation}\label{sssmod}
\begin{split}
\begin{pmatrix}r(\infty) \\ s(\infty)\end{pmatrix}
&= R^{-1}\begin{pmatrix}\rho g(\infty)+am(\infty) \\
\gamma g(\infty) \end{pmatrix}
= R^{-1}\begin{pmatrix}\rho & a \\ \gamma & 0\end{pmatrix}T^{-1}\begin{pmatrix}a \\ 0\end{pmatrix} \\
&= \frac{1}{2[d(a+b)+\rho u][d(d+a+b)+\rho u]}
\begin{pmatrix}2\rho da(d+a)-dau\gamma \\ 2\rho^2a(d+a)+da\gamma(d+a+b)\end{pmatrix}.
\end{split}
\end{equation}

Similarly to the proof in the non-bursty case, the autocorrelation function $C(t)$ has the form of
\begin{equation}\label{correlationmod}
C(t) = (0,1)e^{-Tt}\begin{pmatrix}r(\infty) \\ s(\infty)\end{pmatrix}
+m(\infty)(0,1)(I-e^{-Tt})T^{-1}\begin{pmatrix}a \\ 0\end{pmatrix}-m(\infty)^2.
\end{equation}
Taking $t = 0$ in the above equation and applying Eqs. \eqref{ssm} and \eqref{sssmod} yield
\begin{equation*}
C(0) = s(\infty)-m(\infty)^2
= \frac{2\rho^2da(db+\rho u)+da\gamma(d+a+b)[d(a+b)+\rho u]}{2[d(a+b)+\rho u]^2[d(d+a+b)+\rho u]}.
\end{equation*}
Moreover, taking the derivative on both sides of Eq. \eqref{correlationmod} shows that
\begin{equation*}
C'(t) = -(0,1)e^{-Tt}T\begin{pmatrix}r(\infty) \\ s(\infty)\end{pmatrix}
+m(\infty)(0,1)e^{-Tt}\begin{pmatrix}a \\ 0\end{pmatrix}.
\end{equation*}
Taking $t = 0$ in the above equation yields
\begin{equation*}
C'(0) = -(0,1)T\begin{pmatrix}r(\infty) \\ s(\infty)\end{pmatrix}
= -\frac{da\gamma}{2[d(a+b)+\rho u]}.
\end{equation*}
This shows that the autocorrelation function for the hybrid SDE model has negative derivative at $t = 0$ and thus is monotonically decreasing around $t = 0$. Without loss of generality, we assume that the eigenvalues of the matrix $T$ are distinct. In fact, any matrix can be approximated by such matrices to any degree of accuracy. It then follows from Eq. \eqref{correlationmod} that
\begin{equation*}
C(t) = c_1e^{-\lambda_1t}+c_2e^{-\lambda_2t},
\end{equation*}
where $\lambda_1$ and $\lambda_2$ are the two eigenvalues of the matrix $T$. Note that
\begin{equation}\label{eq1}
c_1+c_2 = C(0) = \frac{2\rho^2da(db+\rho u)+da\gamma(d+a+b)[d(a+b)+\rho u]}{2[d(a+b)+\rho u]^2[d(d+a+b)+\rho u]},
\end{equation}
and we also have
\begin{equation}\label{eq2}
\lambda_1c_1+\lambda_2c_2 = -C'(0) = \frac{da\gamma}{2[d(a+b)+\rho u]}.
\end{equation}
Then $c_1$ and $c_2$ can be calculated by solving the above two equations. This gives the closed-form expression of the autocorrelation function. Taking the Fourier transform gives the following analytical expression of the power spectrum:
\begin{equation*}
G(\omega) = \frac{2c_1\lambda_1}{4\pi^2\omega^2+\lambda_1^2}
+\frac{2c_2\lambda_2}{4\pi^2\omega^2+\lambda_2^2}.
\end{equation*}
This gives Eq. \eqref{positivemod} in the main text.

If the gene is unregulated, then we have $u = 0$ and the two eigenvalues of the matrix $T$ are given by $\lambda_1 = a+b$ and $\lambda_2 = d$. In this case, solving Eqs. \eqref{eq1} and \eqref{eq2} shows that
\begin{equation*}
\begin{split}
&c_1 = \frac{\rho^2ab}{(a+b)^2(d+a+b)(d-a-b))},\\
&c_2 = -\frac{\rho^2ab}{d(a+b)(d+a+b)(d-a-b))}+\frac{a\gamma}{2d(a+b)}.
\end{split}
\end{equation*}
This gives Eq. \eqref{nofeedbackmod} in the main text.

\section*{Acknowledgements}
We thank Ramon Grima, X. Sunney Xie, and Xuejuan Zhang for stimulating discussions and for sharing unpublished results. C. J. acknowledges support from National Natural Science Foundation of China (NSFC) with NSAF grant No. U2230402 and grant No. 12271020. H. Q. thanks partial support from  the Olga Jung Wan Endowed Professorship. M. Q. Z. acknowledges support from NIH with grant No. MH102616 and grant No. MH109665, and also by NSFC with grant No. 31671384 and grant No. 91329000.

\setlength{\bibsep}{5pt}
\footnotesize\bibliographystyle{nature}

\begin{thebibliography}{79}
\expandafter\ifx\csname natexlab\endcsname\relax\def\natexlab#1{#1}\fi
\expandafter\ifx\csname url\endcsname\relax
  \def\url#1{\texttt{#1}}\fi
\expandafter\ifx\csname urlprefix\endcsname\relax\def\urlprefix{URL }\fi

\bibitem[{Murray(2007)}]{murray2007mathematical}
Murray, J.~D.
\newblock \emph{Mathematical Biology: I. An Introduction} (Springer, New York,
  2007), 3rd edn.

\bibitem[{Alon(2006)}]{alon2006introduction}
Alon, U.
\newblock \emph{An Introduction to Systems Biology: Design Principles of
  Biological Circuits} (Chapman and Hall, New York, 2006).

\bibitem[{Qian(2012)}]{qian2012cooperativity}
Qian, H.
\newblock Cooperativity in cellular biochemical processes: {N}oise-enhanced
  sensitivity, fluctuating enzyme, bistability with nonlinear feedback, and
  other mechanisms for sigmoidal responses.
\newblock \emph{Annu. Rev. Biophys.} \textbf{41}, 179--204 (2012).

\bibitem[{Thomas(1981)}]{thomas1981relation}
Thomas, R.
\newblock On the relation between the logical structure of systems and their
  ability to generate multiple steady states or sustained oscillations.
\newblock In \emph{Numerical Methods in the Study of Critical Phenomena},
  180--193 (Springer, 1981).

\bibitem[{Plahte \emph{et~al.}(1995)Plahte, Mestl \&
  Omholt}]{plahte1995feedback}
Plahte, E., Mestl, T. \& Omholt, S.~W.
\newblock Feedback loops, stability and multistationarity in dynamical systems.
\newblock \emph{J. Biol. Syst.} \textbf{3}, 409--413 (1995).

\bibitem[{Gouz{\'e}(1998)}]{gouze1998positive}
Gouz{\'e}, J.-L.
\newblock Positive and negative circuits in dynamical systems.
\newblock \emph{J. Biol. Syst.} \textbf{6}, 11--15 (1998).

\bibitem[{Snoussi(1998)}]{snoussi1998necessary}
Snoussi, E.~H.
\newblock Necessary conditions for multistationarity and stable periodicity.
\newblock \emph{J. Biol. Syst.} \textbf{6}, 3--9 (1998).

\bibitem[{Cinquin \& Demongeot(2002)}]{cinquin2002positive}
Cinquin, O. \& Demongeot, J.
\newblock Positive and negative feedback: striking a balance between necessary
  antagonists.
\newblock \emph{J. Theor. Biol.} \textbf{216}, 229--242 (2002).

\bibitem[{Soul{\'e}(2003)}]{soule2003graphic}
Soul{\'e}, C.
\newblock Graphic requirements for multistationarity.
\newblock \emph{ComPlexUs} \textbf{1}, 123--133 (2003).

\bibitem[{Richard \& Comet(2011)}]{richard2011stable}
Richard, A. \& Comet, J.-P.
\newblock Stable periodicity and negative circuits in differential systems.
\newblock \emph{J. Math. Biol.} \textbf{63}, 593--600 (2011).

\bibitem[{Elowitz \emph{et~al.}(2002)Elowitz, Levine, Siggia \&
  Swain}]{elowitz2002stochastic}
Elowitz, M.~B., Levine, A.~J., Siggia, E.~D. \& Swain, P.~S.
\newblock Stochastic gene expression in a single cell.
\newblock \emph{Science} \textbf{297}, 1183--1186 (2002).

\bibitem[{Golding \emph{et~al.}(2005)Golding, Paulsson, Zawilski \&
  Cox}]{golding2005real}
Golding, I., Paulsson, J., Zawilski, S.~M. \& Cox, E.~C.
\newblock Real-time kinetics of gene activity in individual bacteria.
\newblock \emph{Cell} \textbf{123}, 1025--1036 (2005).

\bibitem[{Cai \emph{et~al.}(2006)Cai, Friedman \& Xie}]{cai2006stochastic}
Cai, L., Friedman, N. \& Xie, X.~S.
\newblock Stochastic protein expression in individual cells at the single
  molecule level.
\newblock \emph{Nature} \textbf{440}, 358--362 (2006).

\bibitem[{Raj \& van Oudenaarden(2008)}]{raj2008nature}
Raj, A. \& van Oudenaarden, A.
\newblock Nature, nurture, or chance: stochastic gene expression and its
  consequences.
\newblock \emph{Cell} \textbf{135}, 216--226 (2008).

\bibitem[{Taniguchi \emph{et~al.}(2010)}]{taniguchi2010quantifying}
Taniguchi, Y. \emph{et~al.}
\newblock Quantifying E. coli proteome and transcriptome with single-molecule
  sensitivity in single cells.
\newblock \emph{Science} \textbf{329}, 533--538 (2010).

\bibitem[{Paulsson(2005)}]{paulsson2005models}
Paulsson, J.
\newblock Models of stochastic gene expression.
\newblock \emph{Phys. Life Rev.} \textbf{2}, 157--175 (2005).

\bibitem[{Schnoerr \emph{et~al.}(2017)Schnoerr, Sanguinetti \&
  Grima}]{schnoerr2017approximation}
Schnoerr, D., Sanguinetti, G. \& Grima, R.
\newblock Approximation and inference methods for stochastic biochemical
  kinetics --- a tutorial review.
\newblock \emph{J. Phys. A: Math. Theor.} \textbf{50}, 093001 (2017).

\bibitem[{Shahrezaei \& Swain(2008)}]{shahrezaei2008analytical}
Shahrezaei, V. \& Swain, P.~S.
\newblock Analytical distributions for stochastic gene expression.
\newblock \emph{Proc. Natl. Acad. Sci. USA} \textbf{105}, 17256--17261 (2008).

\bibitem[{Friedman \emph{et~al.}(2006)Friedman, Cai \&
  Xie}]{friedman2006linking}
Friedman, N., Cai, L. \& Xie, X.~S.
\newblock Linking stochastic dynamics to population distribution: an analytical
  framework of gene expression.
\newblock \emph{Phys. Rev. Lett.} \textbf{97}, 168302 (2006).

\bibitem[{Mackey \emph{et~al.}(2013)Mackey, Tyran-Kaminska \&
  Yvinec}]{mackey2013dynamic}
Mackey, M.~C., Tyran-Kaminska, M. \& Yvinec, R.
\newblock Dynamic behavior of stochastic gene expression models in the presence
  of bursting.
\newblock \emph{SIAM J. Appl. Math.} \textbf{73}, 1830--1852 (2013).

\bibitem[{Bokes \& Singh(2015)}]{bokes2015protein}
Bokes, P. \& Singh, A.
\newblock Protein copy number distributions for a self-regulating gene in the
  presence of decoy binding sites.
\newblock \emph{PloS one} \textbf{10}, e0120555 (2015).

\bibitem[{Jedrak \& Ochab-Marcinek(2016)}]{jkedrak2016time}
Jedrak, J. \& Ochab-Marcinek, A.
\newblock Time-dependent solutions for a stochastic model of gene expression
  with molecule production in the form of a compound Poisson process.
\newblock \emph{Phys. Rev. E} \textbf{94}, 032401 (2016).

\bibitem[{Chen \& Jia(2020)}]{chen2020limit}
Chen, X. \& Jia, C.
\newblock Limit theorems for generalized density-dependent Markov chains and
  bursty stochastic gene regulatory networks.
\newblock \emph{J. Math. Biol.} \textbf{80}, 959--994 (2020).

\bibitem[{Karmakar \& Bose(2004)}]{karmakar2004graded}
Karmakar, R. \& Bose, I.
\newblock Graded and binary responses in stochastic gene expression.
\newblock \emph{Phys. Biol.} \textbf{1}, 197 (2004).

\bibitem[{Raj \emph{et~al.}(2006)Raj, Peskin, Tranchina, Vargas \&
  Tyagi}]{raj2006stochastic}
Raj, A., Peskin, C.~S., Tranchina, D., Vargas, D.~Y. \& Tyagi, S.
\newblock {Stochastic mRNA synthesis in mammalian cells}.
\newblock \emph{PLoS Biol.} \textbf{4}, e309 (2006).

\bibitem[{Dattani \& Barahona(2017)}]{dattani2017stochastic}
Dattani, J. \& Barahona, M.
\newblock Stochastic models of gene transcription with upstream drives: exact
  solution and sample path characterization.
\newblock \emph{J. R. Soc. Interface} \textbf{14}, 20160833 (2017).

\bibitem[{Bressloff(2017)}]{bressloff2017stochastic}
Bressloff, P.~C.
\newblock Stochastic switching in biology: from genotype to phenotype.
\newblock \emph{J. Phys. A: Math. Theor.} \textbf{50}, 133001 (2017).

\bibitem[{Lin \& Buchler(2018)}]{lin2018efficient}
Lin, Y.~T. \& Buchler, N.~E.
\newblock Efficient analysis of stochastic gene dynamics in the non-adiabatic
  regime using piecewise deterministic Markov processes.
\newblock \emph{J. R. Soc. Interface} \textbf{15}, 20170804 (2018).

\bibitem[{Jia \emph{et~al.}(2019)Jia, Yin, Zhang \emph{et~al.}}]{jia2019single}
Jia, C., Yin, G.~G., Zhang, M.~Q. \emph{et~al.}
\newblock Single-cell stochastic gene expression kinetics with coupled
  positive-plus-negative feedback.
\newblock \emph{Phys. Rev. E} \textbf{100}, 052406 (2019).

\bibitem[{Jia \emph{et~al.}(2017)Jia, Zhang \& Qian}]{jia2017emergent}
Jia, C., Zhang, M.~Q. \& Qian, H.
\newblock Emergent Levy behavior in single-cell stochastic gene expression.
\newblock \emph{Phys. Rev. E} \textbf{96}, 040402(R) (2017).

\bibitem[{Wang \emph{et~al.}(2023)Wang, Li \& Jia}]{wang2023poisson}
Wang, X., Li, Y. \& Jia, C.
\newblock Poisson representation: a bridge between discrete and continuous
  models of stochastic gene regulatory networks.
\newblock \emph{J. R. Soc. Interface} \textbf{20}, 20230467 (2023).

\bibitem[{Nov{\'a}k \& Tyson(2008)}]{novak2008design}
Nov{\'a}k, B. \& Tyson, J.~J.
\newblock Design principles of biochemical oscillators.
\newblock \emph{Nat. Rev. Mol. Cell Biol.} \textbf{9}, 981--991 (2008).

\bibitem[{Tsai \emph{et~al.}(2008)}]{tsai2008robust}
Tsai, T. Y.-C. \emph{et~al.}
\newblock Robust, tunable biological oscillations from interlinked positive and
  negative feedback loops.
\newblock \emph{Science} \textbf{321}, 126--129 (2008).

\bibitem[{Li \emph{et~al.}(2017)Li, Liu \& Yang}]{li2017incoherent}
Li, Z., Liu, S. \& Yang, Q.
\newblock Incoherent inputs enhance the robustness of biological oscillators.
\newblock \emph{Cell Syst.} \textbf{5}, 72--81 (2017).

\bibitem[{Lev Bar-Or \emph{et~al.}(2000)}]{lev2000generation}
Lev Bar-Or, R. \emph{et~al.}
\newblock Generation of oscillations by the p53-Mdm2 feedback loop: a
  theoretical and experimental study.
\newblock \emph{Proc. Natl. Acad. Sci. USA} \textbf{97}, 11250--11255 (2000).

\bibitem[{Lahav \emph{et~al.}(2004)}]{lahav2004dynamics}
Lahav, G. \emph{et~al.}
\newblock Dynamics of the p53-Mdm2 feedback loop in individual cells.
\newblock \emph{Nat. Genet.} \textbf{36}, 147 (2004).

\bibitem[{Geva-Zatorsky \emph{et~al.}(2006)}]{geva2006oscillations}
Geva-Zatorsky, N. \emph{et~al.}
\newblock Oscillations and variability in the p53 system.
\newblock \emph{Mol. Syst. Biol.} \textbf{2} (2006).

\bibitem[{Hoffmann \emph{et~al.}(2002)Hoffmann, Levchenko, Scott \&
  Baltimore}]{hoffmann2002ikappab}
Hoffmann, A., Levchenko, A., Scott, M.~L. \& Baltimore, D.
\newblock The I$\kappa$B-NF-$\kappa$B signaling module: temporal control and
  selective gene activation.
\newblock \emph{Science} \textbf{298}, 1241--1245 (2002).

\bibitem[{Nelson \emph{et~al.}(2004)}]{nelson2004oscillations}
Nelson, D. \emph{et~al.}
\newblock Oscillations in NF-$\kappa$B signaling control the dynamics of gene
  expression.
\newblock \emph{Science} \textbf{306}, 704--708 (2004).

\bibitem[{Tay \emph{et~al.}(2010)}]{tay2010single}
Tay, S. \emph{et~al.}
\newblock Single-cell NF-$\kappa$B dynamics reveal digital activation and
  analogue information processing.
\newblock \emph{Nature} \textbf{466}, 267--271 (2010).

\bibitem[{Li \emph{et~al.}(2004)Li, Long, Lu, Ouyang \& Tang}]{li2004yeast}
Li, F., Long, T., Lu, Y., Ouyang, Q. \& Tang, C.
\newblock The yeast cell-cycle network is robustly designed.
\newblock \emph{Proc. Natl. Acad. Sci. USA} \textbf{101}, 4781--4786 (2004).

\bibitem[{Ferrell \emph{et~al.}(2011)Ferrell, Tsai \&
  Yang}]{ferrell2011modeling}
Ferrell, J.~E., Tsai, T. Y.-C. \& Yang, Q.
\newblock Modeling the cell cycle: why do certain circuits oscillate?
\newblock \emph{Cell} \textbf{144}, 874--885 (2011).

\bibitem[{Lee \emph{et~al.}(2000)Lee, Loros \& Dunlap}]{lee2000interconnected}
Lee, K., Loros, J.~J. \& Dunlap, J.~C.
\newblock Interconnected feedback loops in the Neurospora circadian system.
\newblock \emph{Science} \textbf{289}, 107--110 (2000).

\bibitem[{Gallego \& Virshup(2007)}]{gallego2007post}
Gallego, M. \& Virshup, D.~M.
\newblock Post-translational modifications regulate the ticking of the
  circadian clock.
\newblock \emph{Nat. Rev. Mol. Cell Biol.} \textbf{8}, 139--148 (2007).

\bibitem[{Meyer \& Stryer(1988)}]{meyer1988molecular}
Meyer, T. \& Stryer, L.
\newblock Molecular model for receptor-stimulated calcium spiking.
\newblock \emph{Proc. Natl. Acad. Sci. USA} \textbf{85}, 5051--5055 (1988).

\bibitem[{Qi \emph{et~al.}(2020)Qi, Li, Jin, Simmen \&
  Shuai}]{qi2020oscillation}
Qi, H., Li, X., Jin, Z., Simmen, T. \& Shuai, J.
\newblock The oscillation amplitude, not the frequency of cytosolic calcium,
  regulates apoptosis induction.
\newblock \emph{Iscience} \textbf{23}, 101671 (2020).

\bibitem[{Elowitz \& Leibler(2000)}]{elowitz2000synthetic}
Elowitz, M.~B. \& Leibler, S.
\newblock A synthetic oscillatory network of transcriptional regulators.
\newblock \emph{Nature} \textbf{403}, 335 (2000).

\bibitem[{Stricker \emph{et~al.}(2008)}]{stricker2008fast}
Stricker, J. \emph{et~al.}
\newblock A fast, robust and tunable synthetic gene oscillator.
\newblock \emph{Nature} \textbf{456}, 516--519 (2008).

\bibitem[{Barnes \emph{et~al.}(2011)Barnes, Silk \&
  Stumpf}]{barnes2011bayesian}
Barnes, C.~P., Silk, D. \& Stumpf, M.~P.
\newblock Bayesian design strategies for synthetic biology.
\newblock \emph{Interface focus} \textbf{1}, 895--908 (2011).

\bibitem[{Vellela \& Qian(2010)}]{vellela2010}
Vellela, M. \& Qian, H.
\newblock On the Poincar\'{e}-Hill cycle map of rotational random walk:
  locating the stochastic limit cycle in a reversible Schnakenberg model.
\newblock \emph{Proceedings of the Royal Society A: Mathematical, Physical and
  Engineering Sciences} \textbf{466}, 771--788 (2010).

\bibitem[{Qian \& Qian(2000)}]{qian2000pumped}
Qian, H. \& Qian, M.
\newblock Pumped biochemical reactions, nonequilibrium circulation, and
  stochastic resonance.
\newblock \emph{Phys. Rev. Lett.} \textbf{84}, 2271 (2000).

\bibitem[{Bratsun \emph{et~al.}(2005)Bratsun, Volfson, Tsimring \&
  Hasty}]{bratsun2005delay}
Bratsun, D., Volfson, D., Tsimring, L.~S. \& Hasty, J.
\newblock Delay-induced stochastic oscillations in gene regulation.
\newblock \emph{Proc. Natl. Acad. Sci. USA} \textbf{102}, 14593--14598 (2005).

\bibitem[{McKane \emph{et~al.}(2007)McKane, Nagy, Newman \&
  Stefanini}]{mckane2007amplified}
McKane, A.~J., Nagy, J.~D., Newman, T.~J. \& Stefanini, M.~O.
\newblock Amplified biochemical oscillations in cellular systems.
\newblock \emph{J. Stat. Phys.} \textbf{128}, 165--191 (2007).

\bibitem[{Dauxois \emph{et~al.}(2009)Dauxois, Di~Patti, Fanelli \&
  McKane}]{dauxois2009enhanced}
Dauxois, T., Di~Patti, F., Fanelli, D. \& McKane, A.~J.
\newblock Enhanced stochastic oscillations in autocatalytic reactions.
\newblock \emph{Phys. Rev. E} \textbf{79}, 036112 (2009).

\bibitem[{Realpe-Gomez \emph{et~al.}(2012)Realpe-Gomez, Galla \&
  McKane}]{realpe2012demographic}
Realpe-Gomez, J., Galla, T. \& McKane, A.~J.
\newblock Demographic noise and piecewise deterministic Markov processes.
\newblock \emph{Phys. Rev. E} \textbf{86}, 011137 (2012).

\bibitem[{Thomas \emph{et~al.}(2012)Thomas, Straube \& Grima}]{thomas2012slow}
Thomas, P., Straube, A.~V. \& Grima, R.
\newblock The slow-scale linear noise approximation: an accurate, reduced
  stochastic description of biochemical networks under timescale separation
  conditions.
\newblock \emph{BMC Syst. Biol.} \textbf{6}, 1--23 (2012).

\bibitem[{Toner \& Grima(2013)}]{toner2013effects}
Toner, D. \& Grima, R.
\newblock Effects of bursty protein production on the noisy oscillatory
  properties of downstream pathways.
\newblock \emph{Sci. Rep.} \textbf{3}, 2438 (2013).

\bibitem[{Thomas \emph{et~al.}(2013)Thomas, Straube, Timmer, Fleck \&
  Grima}]{thomas2013signatures}
Thomas, P., Straube, A.~V., Timmer, J., Fleck, C. \& Grima, R.
\newblock Signatures of nonlinearity in single cell noise-induced oscillations.
\newblock \emph{J. Theor. Biol.} \textbf{335}, 222--234 (2013).

\bibitem[{Thomas \emph{et~al.}(2014)Thomas, Popovi{\'c} \&
  Grima}]{thomas2014phenotypic}
Thomas, P., Popovi{\'c}, N. \& Grima, R.
\newblock Phenotypic switching in gene regulatory networks.
\newblock \emph{Proc. Natl. Acad. Sci. USA} \textbf{111}, 6994--6999 (2014).

\bibitem[{Jia \& Grima(2021)}]{jia2021frequency}
Jia, C. \& Grima, R.
\newblock Frequency domain analysis of fluctuations of mRNA and protein copy
  numbers within a cell lineage: theory and experimental validation.
\newblock \emph{Physical Review X} \textbf{11}, 021032 (2021).

\bibitem[{Jia \emph{et~al.}(2022)Jia, Singh \& Grima}]{jia2022concentration}
Jia, C., Singh, A. \& Grima, R.
\newblock Concentration fluctuations in growing and dividing cells: Insights
  into the emergence of concentration homeostasis.
\newblock \emph{PLoS Comput. Biol.} \textbf{18}, e1010574 (2022).

\bibitem[{Gupta \& Khammash(2022)}]{gupta2022frequency}
Gupta, A. \& Khammash, M.
\newblock Frequency spectra and the color of cellular noise.
\newblock \emph{Nat. Commun.} \textbf{13}, 4305 (2022).

\bibitem[{Lestas \emph{et~al.}(2010)Lestas, Vinnicombe \&
  Paulsson}]{lestas2010fundamental}
Lestas, I., Vinnicombe, G. \& Paulsson, J.
\newblock Fundamental limits on the suppression of molecular fluctuations.
\newblock \emph{Nature} \textbf{467}, 174--178 (2010).

\bibitem[{Ashall \emph{et~al.}(2009)}]{ashall2009pulsatile}
Ashall, L. \emph{et~al.}
\newblock Pulsatile stimulation determines timing and specificity of
  NF-$\kappa$B-dependent transcription.
\newblock \emph{Science} \textbf{324}, 242--246 (2009).

\bibitem[{Hornos \emph{et~al.}(2005)}]{hornos2005self}
Hornos, J. \emph{et~al.}
\newblock Self-regulating gene: an exact solution.
\newblock \emph{Phys. Rev. E} \textbf{72}, 051907 (2005).

\bibitem[{Grima \emph{et~al.}(2012)Grima, Schmidt \& Newman}]{grima2012steady}
Grima, R., Schmidt, D. \& Newman, T.
\newblock Steady-state fluctuations of a genetic feedback loop: An exact
  solution.
\newblock \emph{J. Chem. Phys.} \textbf{137}, 035104 (2012).

\bibitem[{Kumar \emph{et~al.}(2014)Kumar, Platini \& Kulkarni}]{kumar2014exact}
Kumar, N., Platini, T. \& Kulkarni, R.~V.
\newblock Exact distributions for stochastic gene expression models with
  bursting and feedback.
\newblock \emph{Phys. Rev. Lett.} \textbf{113}, 268105 (2014).

\bibitem[{Jia \& Grima(2020{\natexlab{a}})}]{jia2020small}
Jia, C. \& Grima, R.
\newblock Small protein number effects in stochastic models of autoregulated
  bursty gene expression.
\newblock \emph{J. Chem. Phys.} \textbf{152}, 084115 (2020{\natexlab{a}}).

\bibitem[{Jia \& Grima(2020{\natexlab{b}})}]{jia2020dynamical}
Jia, C. \& Grima, R.
\newblock Dynamical phase diagram of an auto-regulating gene in fast switching
  conditions.
\newblock \emph{J. Chem. Phys.} \textbf{152}, 174110 (2020{\natexlab{b}}).

\bibitem[{Jiao \emph{et~al.}(2015)Jiao, Sun, Tang, Yu \&
  Zheng}]{jiao2015distribution}
Jiao, F., Sun, Q., Tang, M., Yu, J. \& Zheng, B.
\newblock Distribution modes and their corresponding parameter regions in
  stochastic gene transcription.
\newblock \emph{SIAM J. Appl. Math.} \textbf{75}, 2396--2420 (2015).

\bibitem[{Cai \emph{et~al.}(2008)Cai, Dalal \& Elowitz}]{cai2008frequency}
Cai, L., Dalal, C.~K. \& Elowitz, M.~B.
\newblock Frequency-modulated nuclear localization bursts coordinate gene
  regulation.
\newblock \emph{Nature} \textbf{455}, 485 (2008).

\bibitem[{Qian \emph{et~al.}(2003)Qian, Qian \& Zhang}]{qian2003fundamental}
Qian, M., Qian, M.-P. \& Zhang, X.-J.
\newblock Fundamental facts concerning reversible master equations.
\newblock \emph{Phys. Lett. A} \textbf{309}, 371--376 (2003).

\bibitem[{Jia \& Chen(2015)}]{jia2015second}
Jia, C. \& Chen, Y.
\newblock A second perspective on the Amann--Schmiedl--Seifert criterion for
  non-equilibrium in a three-state system.
\newblock \emph{J. Phys. A: Math. Theor.} \textbf{48}, 205001 (2015).

\bibitem[{Suter \emph{et~al.}(2011)}]{suter2011mammalian}
Suter, D.~M. \emph{et~al.}
\newblock Mammalian genes are transcribed with widely different bursting
  kinetics.
\newblock \emph{Science} \textbf{332}, 472--474 (2011).

\bibitem[{Ginoux \& Letellier(2012)}]{ginoux2012van}
Ginoux, J.-M. \& Letellier, C.
\newblock Van der Pol and the history of relaxation oscillations: Toward the
  emergence of a concept.
\newblock \emph{Chaos} \textbf{22} (2012).

\bibitem[{Zenklusen \emph{et~al.}(2008)Zenklusen, Larson \&
  Singer}]{zenklusen2008single}
Zenklusen, D., Larson, D.~R. \& Singer, R.~H.
\newblock Single-RNA counting reveals alternative modes of gene expression in
  yeast.
\newblock \emph{Nat. Struct. Mol. Biol.} \textbf{15}, 1263 (2008).

\bibitem[{Jiao \& Zhu(2020)}]{jiao2020regulation}
Jiao, F. \& Zhu, C.
\newblock Regulation of gene activation by competitive cross talking pathways.
\newblock \emph{Biophys. J.} \textbf{119}, 1204--1214 (2020).

\bibitem[{Chen \emph{et~al.}(2022)Chen, Lin \& Jiao}]{chen2022using}
Chen, L., Lin, G. \& Jiao, F.
\newblock Using average transcription level to understand the regulation of
  stochastic gene activation.
\newblock \emph{R. Soc. Open Sci.} \textbf{9}, 211757 (2022).

\bibitem[{Jia \& Li(2023)}]{jia2023analytical}
Jia, C. \& Li, Y.
\newblock Analytical time-dependent distributions for gene expression models
  with complex promoter switching mechanisms.
\newblock \emph{SIAM J. Appl. Math.} \textbf{83}, 1572--1602 (2023).

\end{thebibliography}

\end{document}